\documentclass[12pt,12pt]{article}
\usepackage{graphicx}
\usepackage{epsfig}

\catcode`\@=11
\newdimen\ex@
\font\dozeb=cmmib10 scaled \magstep1
\font\dozesyb=cmbsy10 scaled \magstep1
\font\dezb=cmmib10
\def\beq{\begin{equation}}
\def\eeq{\end{equation}}
\def\beqa{\begin{eqnarray}}
\def\eeqa{\end{eqnarray}}

\def\b{{\beta}}

\begin{document}
\def\thefootnote{\fnsymbol{footnote}}

\title{
\bf
Resonating mean-field theoretical approach \\
to two-gap superconductivity with high-$T_c$\footnotemark[1]
}
\author{
Seiya Nishiyama$\!$\footnotemark[2] $\!\!$,
Jo\~ao da Provid\^{e}ncia$\!$\footnotemark[3]\\
Constan\c{c}a Provid\^{e}ncia\footnotemark[4]~~and
Hiromasa Ohnishi\footnotemark[5]~\\
\\
Centro de F\'\i sica Computacional,
Departamento de F\'\i sica,\\
Universidade de Coimbra,
3000-Coimbra, $\!$Portugal\footnotemark[2] \\
\\
Electronic Materials Simulation Research Group,\\
Research Initiative of Computational Science (RICS),\\ 
Nanosystem Research Institute (NRI),\\ 
National Institute of Advanced Science and Technology (AIST),\\
1-1-1, Umezono, Tsukuba, Ibaraki 305-8568, Japan\footnotemark[5]\\
\\[-4pt]
{\it Dedicated to the Memory of Hideo Fukutome}
}

\maketitle

\footnotetext[1]{The original work has been presented by
Nishiyama, da Provid\^encia and Ohnishi,\\
~~~~~~in {\it the 26th and 29th International Workshop on
Condensed Matter Theories}
\cite{NPOProceedings.0307}.}
\footnotetext[2]{
Corresponding author. E-mail address: seikoceu@khe.biglobe.ne.jp}
\footnotetext[3]{
E-mail address: providencia@teor.fis.uc.pt}
\footnotetext[4]
{E-mail address: cp@teor.fis.uc.pt}
\footnotetext[5]
{E-mail address: ohni@aist.go.jp}

\vspace{-0.75cm}


\begin{abstract}
\vskip0.1cm
In the recent paper (referred to as I),
the resonating mean-field theory (Res-MFT)
has been applied and shown to effectively describe
the so-called two-gap superconductivity (SC).
In I, a state with large quantum fluctuations
has been approximated by superposition of two MF wave functions
composed of Hartree-Bogoliubov (HB) wave functions 
with different correlation structures.
Particularly in I,
using a suitable chemical potential,
at $T \!=\! 0$ the two-gap SC in MgB$_2$ has been well described by the Res-HBT.
Furthermore
the Res-HB ground state generated with HB wave functions
has almost explained value of the ground-state correlation energy
in all the correlation regimes
including an intermediate coupling regime.
In the present paper
we will apply the Res-HBT to
the two-gap SC with {\em high} critical temperature $T_c$.
We will aim at constructing a theoretical foundation
for phenomenological theories of the two-gap SC
at $T \!=\! 0$ and finite temperature.
In the single-gap case
we will find a new formula leading to a higher $T_c$ than
the usual HB's.

\vspace{0.2cm}

\end{abstract}

\newpage


\def\thesection{\arabic{section}}
\setcounter{equation}{0}
\renewcommand{\theequation}{\arabic{section}.\arabic{equation}}

\section{Introduction}

~~~~A topical two-gap superconductivity (SC)
with critical temperature $T_c \!=\! 39$K
has been recently discovered
in $\mbox{MgB}_2$
\cite{Akimitsu.01}.
It may be expected
to open a new area in the vigorous pursuit by the radical spirit of
the resonating mean-field theories (Res-MFTs)
\cite{Fuku.88,NishiFuku.91}
to develop a theoretical framework appropriate to
explore the problem of {\em high} $T_c$ in superconductors.
In particular,
fermion systems with large quantum fluctuations
show serious difficulties in many-body problems at finite temperature.
To approach such problems,
Fukutome has developed the resonating
Hartree-Fock theory (Res-HFT)
\cite{Fuku.88}
and Fukutome and one of the present authors (S.N.)
have extended it directly to
the resonating Hartree-Bogoliubov theory (Res-HBT)
to include pair correlations
\cite{NishiFuku.91,NishiFuku.92}.
In the recent paper (referred to as I)
\cite{NPO.06},
the Res-HBT has been applied
and demonstrated to effectively describe one of an exciting topics of current interest, the two-gap SC.
An appearance of {\it Thermal Gap Equation} in the Res-HBT
is a manifestation of the analogy of
the Res-HBT with the usual BCS and HBT
\cite{BCS.57,Bogo.59}.

Before the discovery of {\it high}-$T_c$ superconductor,
much effort had been devoted to raising
$T_c$
of the usual BCS superconductor in the weak coupling regime
\cite{BCS.57,Bogo.59,BTS.59}
and to obtaining the Eliashberg's critical temperature
in the strong coupling regime
\cite{Eliashberg.60,Schrieffer.64,GK.82}.
The $T_c$ for
$\mbox{MgB}_2$
is 39K, which
is close to or even higher than the upper theoretical value
predicted by the BCS theory
\cite{Parks.69}.
Even if SC in
$\mbox{MgB}_2$
is phonon-mediated,
a model beyond the simple BCS model or the Eliashberg model
is required.
The existence of
two energy-gaps in
$\mbox{MgB}_2$
at $T\!=\!0$ has been predicted phenomenologically by Kortus et al.
\cite{KMBAB.01}
and Liu et al.
\cite{LMK.01}
employing a BCS-like weak-coupling theory,
using the effective $\sigma$ and $\pi$ two-band model.
They have obtained
$\Delta _{\sigma } \!=\! 7.4$ [meV]
and
$\Delta _{\pi } \!=\! 2.4$ [meV].
On the other hand,
employing the Eliashberg's strong-coupling theory
\cite{Eliashberg.60},
Choi et al.
\cite{CRSCL.02}
have obtained
$\Delta _{\sigma } \!=\! 6.8$ [meV] 
and 
$\Delta _{\pi } \!=\! 1.8$ [meV].
Beyond such theoretical great successes,
the Res-HB ground state generated with HB wave functions
which are equivalent to the coherent state representations (CS reps)
\cite{Perelomov.72},
is expected to almost explain
the value of the ground-state correlation energy
in all the correlation regimes
including an intermediate coupling regime.

To go beyond such phenomenologies,
we develop a tentative temperature-dependent Res-HBT.
A temperature dependent variation is made
to satisfy diagonalization conditions for
thermal Res-Fock-Bogoliubov (Res-FB) operators
along a way different from the usual thermal-BCS theory
\cite{KA.59,AGD.65,Abrikosov.88}.
We derive formulas for determining
$T_c$ and the behaviour of the gap near $T \!=\! 0$ and $T_c$.
In the particular case of an {\em equal} magnitude of gaps but with two different phases,
we find new formulas boosting up $T_c$ to a higher value than
the usual HB's value.
Finally we stress to attempt a projection-method approach to
a rigurous thermal Res-HBT.
A partition function in an $SO(2N)$ ($N$: Number of fermion states) group
can be calculated using the projection method proposed in the Res-MFTs
\cite{Fuku.88,NishiFuku.91}

In Section 2,
we give a diagonalization condition for
thermal Res-FB operators
which is the genralization of the condition to the Res-HBT's
proposed by the Ozaki's method
\cite{Ozaki.85}.
The diagonalized thermal Res-HB density matrix
is expressed in the form of
a Fermi-Dirac distribution with Res-HB eigenvalues.
These lead to a self-consistent
thermal Res-HB gap equation,
from which we get formulas to determine $T_c$ and
the behaviour of gaps near $T \!=\! 0$ and $T_c$.
Finally in Section 3,
we give a summary and further perspectives.
In Appendices,
we give a resonating mean-field free energy,
variation of the resonating mean-field free energy,
diagonalization of thermal HB density matrix and
necessary integral formulas
to calculate new formulas for the gap at intermediate temperature.


\def\thesection{\arabic{section}}
\setcounter{equation}{0}
\renewcommand{\theequation}{\arabic{section}.\arabic{equation}}

\section{Thermal resonating HB equation}

~~
The Res-HB eigenvalue equation
${\cal F}_r u_{rp} \!=\! \epsilon _{rp}u_{rp}$
in I can be extended to the thermal Res-HB eigenvalue equation.
We obtain the thermal Res-HB equation in Appendix A.
The Res-HBT implies that
every HB eigenfunction in an HB resonating state
has its own orbital-energy.
We can derive the thermal Res-HB eigenvalue equation
which is given in Appendix B.
The thermal HB density matrix is determined as
$
W_{rrp}[{\cal F}_{rp}]
\! = \!
\left[
1 \!+\! \exp \{ \beta {\cal F}_{rp} \}
\right]^{-1} \!
$
which is proved also in Appendix B.
Using a Bogoliubov transformation $g_{1p}$ and $g_{2p}$ in I
and
(\ref{tildeCalFrmatrix}) and (\ref{CalFrmatrix}),
$W_{11p}[{\cal F}_{1p} ]$ and $W_{22p}[{\cal F}_{2p} ]$ in I
are diagonalized as follows:\\[-18pt]
\beqa
\widetilde{W}_{rp}
\!=\!
g_{rp} ^{\dagger } W_{rrp}[{\cal F}_{rp} ] g_{rp}
\!=\!
\left[ \!\!
\begin{array}{cc}
\widetilde{w}_{rp} &\!\! 0\\
\\[-10pt]
0 &\!\! 1 - \widetilde{w}_{rp}
\end{array} \!\!
\right] ,
~(r\!=\!1,2) ,
\label{tildeCalFrmatrix0}
\eeqa
\vspace{-0.7cm}
\beqa
\widetilde{w}_{rp}
\!=\!
{\displaystyle
\frac{1}{1 \!+\!
e^{{\displaystyle \beta \widetilde{\epsilon }_{rp}}}}
},~~
1 \!-\! \widetilde{w}_{rp}
\!=\!
{\displaystyle
\frac{1}{1 \!+\!
e^{{\displaystyle -\beta \widetilde{\epsilon }_{rp}}}}
} ,~~
\left( \!
\beta
\! = \!
{\displaystyle
\frac{1}{k_B T}
} \!
\right) ,
\label{tildethermaldensitymat0}
\eeqa\\[-12pt]
which are the generalizations of the Ozaki's results
\cite{Ozaki.85}
to the Res-MFT.
By making the Bogoliubov transformation $g_{rp}$,
eigenvalues
$\widetilde{\epsilon }_{rp}$
are obtained by diagonalization of
the thermal Res-FB operators ${\cal F}_{rp}$
with additional terms
$( H[W_{rrp}] \!-\! E ) |c_{rp}|^2~(r\!=\!1,2)$
(\ref{CalFrmatrix}).
The thermal HB interstate density matrix
in the whole Res-HB subspace
is given as the direct sum:\\[-18pt]
\beqa
\begin{array}{l}
W_p [{\cal F}_p]
\!=\!
g_p \widetilde{W_p}g_p ^{\dagger }
\!=\!
\bigoplus_{r=1}^2 W_{rrp}[{\cal F}_{rp}],~~
W_{rr}[{\cal F}_r ]
\!=\!
g_{rp} \widetilde{W}_{rp} g_{rp} ^{\dagger } .
\end{array}
\label{WrrCalFrmatrix}
\eeqa\\[-30pt]

Suppose a {\em tilde} thermal Res-HB density operator
$\widetilde{W}_{1p}$
for {\em equal-gaps} to be\\[-18pt]
\beqa
\widetilde{W}_{1p}
\!=\!
\left[ \!\!
\begin{array}{cc}
\widetilde{W}_{1p} ^{\uparrow } \!\cdot\! I_2 & 0 \\
\\[-10pt]
0 & \widetilde{W}_{1p} ^{\downarrow } \!\cdot\! I_2
\end{array} \!\!
\right] ,~~
\widetilde{W}_{1p} ^{\uparrow (\downarrow)}
\!=\!
\left[ \!\!
\begin{array}{cc}
\widetilde{w}_{1p} ^{\uparrow (\downarrow)} \!\cdot\! I_2 & 0 \\
\\[-10pt]
0 & (1 \!-\! \widetilde{w}_{1p} ^{\uparrow (\downarrow)})
\!\cdot\! I_2
\end{array} \!\!
\right] .
\\[-12pt] \nonumber
\eeqa
Here $I_2$ is the two-dimensional unit matrix.
Performing the unitary transformation by
$\widehat{g}_{1p} ^{\uparrow (\downarrow)}$,
we obtain the following thermal Res-HB density matrix
$\widetilde{W}_{1p} ^{\uparrow (\downarrow)}$.\\[-18pt]
\beqa
\begin{array}{rl}
&\!\!\!\!\!\!\!\!
W_{1p}^{\uparrow (\downarrow)}
\!=\!
\widehat{g}_{1p} ^{\uparrow (\downarrow)}
\widetilde{W}_{1p} ^{\uparrow (\downarrow)}
\widehat{g}_{1p} ^{\uparrow (\downarrow)\dagger }
\!=\!
\widehat{g}_{1p} ^{\uparrow (\downarrow)} \!
\left[ \!\!
\begin{array}{cc}
\widetilde{w}_{1p} ^{\uparrow (\downarrow)} \!\cdot\! I_2 & 0 \\
\\[-10pt]
0 & (1 \!-\! \widetilde{w}_{1p} ^{\uparrow (\downarrow)})
\! \!\cdot\! I_2 \!
\end{array} \!\!
\right] \!
\widehat{g}_{1p} ^{\uparrow (\downarrow)\dagger } \\
\\[-6pt]
&\!\!\!\!\!\!\!\!
\!=\!\!
\left[ \!\!\!
\begin{array}{cc}
\frac{1}{2} \!
\left\{ \!
1 \!\!-\!\!
\cos \widehat{\theta } _{1p}
\left( \!
1 \!\!-\!\! 2 \widetilde{w}_{1p} ^{\uparrow (\downarrow)} \!
\right) \!
\right\}
\! \!\cdot\! I_2 \!
&\!\!\!\!\!\!
\{ \!-(+)\! \}  \! \times \!
\frac{1}{2} \!
\sin \widehat{\theta } _{1p}e ^{-i\widehat{\psi } _1} \!
\left( \!
1 \!\!-\!\! 2 \widetilde{w}_{1p} ^{\uparrow (\downarrow)}
\right)
\! \!\cdot\! I_2 \!
\\[-6pt]
&\\
\{ \!-(+)\! \} \! \times \!
\frac{1}{2} \!
\sin \widehat{\theta } _{1p}e ^{i\widehat{\psi } _1} \!
\left( \!
1 \!\!-\!\! 2 \widetilde{w}_{1p} ^{\uparrow (\downarrow)}
\right)
\! \!\cdot\! I_2 \!
&\!\!\!\!\!\!
\frac{1}{2} \!
\left\{ \!
1 \!\!+\!\!
\cos \widehat{\theta } _{1p} \!
\left( \!
1 \!\!-\!\! 2 \widetilde{w}_{1p} ^{\uparrow (\downarrow)} \!
\right) \!
\right\}
\! \!\cdot\! I_2 \!
\\
\end{array}
\!\!
\right] \! .
\end{array}
\label{Wrrmatrix2}
\eeqa
The
$\widetilde{W}_{2p} ^{\uparrow (\downarrow)}$
has the same form as
(\ref{Wrrmatrix2}).
The Res-FB operator
${\cal F}^{\uparrow }_{1(2)p}$
for spin-up state
with upper and lower signs,
corresponding to Case I
(4.5)
and Case II
(4.6) in I,
is expressed as\\[-16pt]
\beqa
{\cal F}^{\uparrow }_{1(2)p}
\!=\!
\left[ \!\!
\begin{array}{cc}
{\cal F}^{\uparrow }_{+ \varepsilon_p} \!\cdot\! I_2
&\{+ (-)\} \!\times\! {\cal F}^{\uparrow }_{\Delta p} \!\cdot\! I_2 \\
\\[-8pt]
\{+ (-)\} \!\times\! {\cal F}^{\uparrow }_{\Delta p} \!\cdot\! I_2
&-{\cal F}^{\uparrow }_{- \varepsilon_p} \!\cdot\! I_2
\end{array} \!\!
\right] .
\label{ResFB1p}
\eeqa
\vspace{-0.5cm}
\beqa
\!\!\!\!\!\!\!\!\!\!
\begin{array}{c}
{\cal F}^{\uparrow }_{\!+(-) \varepsilon _p} 
\!\!\equiv\!\!
{\displaystyle \frac{1}{2}} \!\!
\left\{ \!
\varepsilon_p
\!\!+\!\!
2
(\!
H[W] \!\!-\!\!
E_{\mbox{\scriptsize gr}}^{\mbox{\scriptsize Res}}
\!)
{\displaystyle 
\frac
{
\sin ^{2} \!
\frac{\theta _p}{2}
\!
\left( \!
\cos ^{2} \!
\frac{\theta _p}{2} \!
\right)
}
{\cos {\theta }_p}
\!\mp\!
\frac{\Delta ^2}{\varepsilon_p}}
\!\cdot\! [\det \! z_{12}]^{\frac{1}{2}} \!
\right\}
\!\cdot\!
{\displaystyle \frac{1}
{1 \!\!\pm\!\! [\det \! z_{12}]^{\frac{1}{2}}}} , 
\end{array} \!\!\!
\label{ResFepandFdp1}
\eeqa
\vspace{-0.5cm}
\beqa
\!\!\!\!\!\!\!\!\!\!
\begin{array}{c}
{\cal F}^{\uparrow }_{\!\Delta p} 
\!\!=\!\!
{\cal F}^{\uparrow }_{\!\Delta }
\!\!\equiv\!\!
-{\displaystyle
\frac{1}{2}} \!
\Delta \!
\left\{ \!
N(0) V
\!\cdot\! \mbox{arcsinh}
\left( \!
{\displaystyle \frac{1}{x}} \!
\right)
\!\pm\!
[\det \! z_{12}]^{\frac{1}{2}} \!
\right\}
\!\cdot\!
{\displaystyle \frac{1}
{1 \!\!\pm\!\! [\det \! z_{12}]^{\frac{1}{2}}}},
\left(\!
x
\!=\!
{\displaystyle \frac{\Delta }{\hbar \omega_D}} \!
\right) ,
\end{array} \!\!\!
\label{ResFepandFdp2}
\eeqa
for Case I(upper sign) and Case II(lower sign) where
\beqa
\!\!\!\!\!\!\!\!\!\!
\begin{array}{c}
[\det{z}_{12}]^{\frac{1}{2}} 
\!=\!
\exp
\left[
- 2 N(0) \hbar \omega_D
\left\{
{\displaystyle
\ln (1 \!+\! x^2 ) \!+\! 2 x
\!\cdot\! \arctan \!
\left( \!
\frac{1}{x} \!
\right)
} \!
\right\}
\right] .
\end{array} \!\!\!
\label{detz}
\eeqa
At finite temperature,
using the formulas
(\ref{tildeCalFrmatrix0})
and
(\ref{tildethermaldensitymat0})
we require orrespondence relations
$\cos \theta _{pT}
\!\Rightarrow\!
\cos \widehat{\theta } _{1p}$
and
$\sin \theta _{pT}
\!\Rightarrow\!
\sin \widehat{\theta } _{1p}$
given through\\[-16pt]
\beqa
\!\!\!\!
\left.
\begin{array}{c}
{\displaystyle
\cos \theta _{p}
\!\!=\!\!
\frac{\varepsilon _p}
{\sqrt{\varepsilon _p ^2 \!\!+\!\! \Delta ^2 _{\!T}}}
\!=\!
\frac{
{\cal F}^{\uparrow }_{\!\! + \varepsilon_p T}
\!\!+\!\!
{\cal F}^{\uparrow }_{\!\! - \varepsilon_p T} \!
}
{2 ~\! \widetilde{\epsilon }_{1p} ^{\uparrow }} \!
\left( \!
1 \!\!-\!\! 2 \widetilde{w}_{1p} ^{\uparrow } \!
\right) 
} ,
\\
\\[-14pt]
{\displaystyle
\sin \theta _{p}
\!\!=\!\!
\frac{\Delta _{\!T}}
{\sqrt{\varepsilon _p ^2 \!\!+\!\! \Delta ^2 _{\!T}}}
\!=\!
- \frac{{\cal F}_{\!\! \Delta _T} ^{\uparrow }}
{\widetilde{\epsilon }_{1p} ^{\uparrow }} \!
\left( \!
1 \!\!-\!\! 2 \widetilde{w}_{1p} ^{\uparrow } \!
\right) ,
}
\end{array} \!\!\!
\right\}
\!\rightarrow\!
\frac{\Delta _{\!T}}{\varepsilon _p }
\!\!=\!\!
-
\frac{{\cal F}_{\!\! \Delta _T} ^{\uparrow } \!\!
\left( \!
1 \!\!-\!\! 2 \widetilde{w}_{1p} ^{\uparrow } \!
\right)
}
{\displaystyle
{\frac{{\cal F}^{\uparrow }_{\!\! + \varepsilon_p T}
\!\!+\!\!
{\cal F}^{\uparrow }_{\!\! - \varepsilon_p T}}{2}
} \!
\left( \!
1 \!\!-\!\! 2 \widetilde{w}_{1p} ^{\uparrow } \!
\right)
} .
\label{equivalenceforcosandsin}
\eeqa\\[-10pt]
Notice the multiplication factor
$
1 \!\!-\!\! 2 \widetilde{w}_{1p} ^{\uparrow (\downarrow)}
$.
The
$
\widetilde{\epsilon }^{\uparrow }_{1p}
(
\!=\!
\widetilde{\epsilon }^{\uparrow }_p
)
$
is the quasi-particle (QP) energy:
$
\widetilde{\epsilon }^{\uparrow }_p
\!\!=\!\!
\sqrt{
\left( \!
{\cal F}^{\uparrow }_{\! + \varepsilon_p T}
\!+\!
{\cal F}^{\uparrow }_{\! - \varepsilon_p T} \!
\right)^2 \!\!
/ 4
\!+\!
{\cal F}_{\! \Delta _T} ^{\uparrow 2} \!
}
$.
Equations in L.H.S. of
(\ref{equivalenceforcosandsin})
are unified into one equation in R.H.S.
It is easily shown that equation
(\ref{equivalenceforcosandsin})
plays the role of
self-consistency condition
at $T = 0$.
Dividing numerator and denominator, respectively
by
$(\varepsilon _p ^2 + \Delta ^2 _T) ^{3/2}$,
equation in R.H.S. of
(\ref{equivalenceforcosandsin})
is rewritten as\\[-16pt]
\beqa
1
\!=\!
\frac{
{\displaystyle
\frac{\varepsilon _p ^2}
{(\varepsilon _p ^2 \!+\! \Delta ^2 _T) ^{\frac{3}{2}}} \!
\left( \!
- \frac{2{\cal F}_{\Delta _T} ^{\uparrow }}{\Delta _T} \!
\right)
} \!
\left( \!
1 \!-\! 2 \widetilde{w}_{p} ^{\uparrow }
\right)
}
{
{\displaystyle
\frac{\varepsilon _p}
{(\varepsilon _p ^2 + \Delta ^2 _T) ^{\frac{3}{2}}}
} \!
\left( \!
{\cal F}^{\uparrow }_{+ \varepsilon_p T}
\!+\!
{\cal F}^{\uparrow }_{- \varepsilon_p T} \!
\right) \!
\left( \!
1 \!-\! 2 \widetilde{w}_{p} ^{\uparrow } \!
\right)
} ,~
\widetilde{w}_{p} ^{\uparrow }
\!=\!
\widetilde{w}_{1p} ^{\uparrow } .
\label{equivalenceforsin3}
\eeqa\\[-10pt]
Now we demand a new condition for
thermal {\it gap equation}\\[-16pt]
\beqa
\begin{array}{c}
\sum_p \!
\left\{ \!
{\displaystyle
\frac{\varepsilon _p}
{
(\varepsilon _p ^2 \!+\! \Delta ^2 _T) ^{\frac{3}{2}}
}
{\displaystyle \frac{1}{2}} \!
\left( \!
{\cal F}^{\uparrow }_{+ \varepsilon_p T}
\!+\!
{\cal F}^{\uparrow }_{ - \varepsilon_p T} \!
\right)
}
\!-\!
{\displaystyle
\frac{\varepsilon _p ^2}
{
(\varepsilon _p ^2 \!+\! \Delta ^2 _T) ^{\frac{3}{2}}
} \!
\left( \!
- \frac{{\cal F}_{\Delta _T} ^{\uparrow }}{\Delta _T} \!
\right)
} \!
\right\} \!
\left(
1 \!-\! 2 \widetilde{w}_{p} ^{\uparrow }
\right)
\!=\! 0 ,
\end{array}
\label{gapequation}
\eeqa
which leads to\\[-20pt]
\beqa
\left.
\begin{array}{c}
\left\{ \!
1 \!-\! N(0) V
\!\cdot\! \mbox{arcsinh} \!
\left( \!
{\displaystyle \frac{1}{x_T}} \!
\right)
\!\mp\!
[\det z_{12}]_T ^{\frac{1}{2}} \!
\right\} \!
\sum _p \! A_p  \\
\\[-14pt]
+
\widetilde{E}_{\mbox{\scriptsize gr}_T}^{\mbox{\scriptsize Res}(\pm)}
\hbar \omega_D \!
\sum _p \! B_p
\!\mp\!
\Delta ^2 _T \!\cdot\! [\det z_{12}]_T ^{\frac{1}{2}}
\sum _p \! C_p
\!=\! 0 , \\
\\[-10pt]
\!\!\!\!\!\!\!\!
{\displaystyle
\frac{
\widetilde{E}_{\mbox{\scriptsize gr}_T}^{\mbox{\scriptsize Res}(\!\pm\!)}}
{\hbar \omega_D}
\!\!=\!\!
\pm 2 \!
N \! (0) \hbar \omega_D x_{\!T} ^2
\!\cdot\! \mbox{arcsinh} \!
\left( \!
\frac{1}{x_{\!T}} \!
\right) \!\!
\left\{
2 \!\! - \!\! N \! (0) \! V \!
\!\cdot\! \mbox{arcsinh} \!
\left( \!
\frac{1}{x_{\!T}} \!
\right) \!
\right\} \!
\frac{[\det z_{12}]_T ^{\frac{1}{2}}}{1 
\!\!\pm\!\! 
[\det z_{12}]_T ^{\frac{1}{2}}}
} .
\end{array} \!\!\!
\right\}
\label{TempSCFcondition}
\eeqa
Here we calculate the term 
$(H[W] \!-\! E_{\mbox{\scriptsize gr}}^{\mbox{\scriptsize Res}})$
in
(\ref{ResFepandFdp1})
using the solutions for the Res-HB CI equation obtained in I
and denote the result as
$
\widetilde{E}_{\mbox{\scriptsize gr}T}^{\mbox{\scriptsize Res}(\!\pm\!)}
$.
We also give the following definitions for
$\sum _p \! A_p,~\sum _p \! B_p$ and $\sum _p \! C_p$:\\[-20pt]
\beqa
\!\!\!\!
\begin{array}{c}
\left[ \!
\sum _p \! A_p,\sum _p \! B_p,\sum _p \! C_p \!
\right]
\!\equiv\!
\sum _p \!
{\displaystyle
\left[ \!
\frac
{\varepsilon _p ^2}
{(\varepsilon _p ^2 \!+\! \Delta ^2 _T) ^{\frac{3}{2}}},
\frac
{1}
{\varepsilon _p ^2 \!+\! \Delta ^2 _T},
\frac
{1}
{(\varepsilon _p ^2 \!+\! \Delta ^2 _T) ^{\frac{3}{2}}} \!
\right] \!
\left( \!
1 \! - \! 2 \widetilde{w}_{p} ^{\uparrow } \!
\right)
}.
\end{array}
\label{DefinitionsofABandC}
\eeqa\\[-12pt]
Rearranging 
(\ref{TempSCFcondition}),
it is cast to\\[-22pt]
\beqa
&\!\!\!\!\!\!\!\!\!\!\!\!\!
{\displaystyle
\frac{1}{N(0) V}
}
\!=\!
{\displaystyle
\mbox{arcsinh} \!
\left( \!
\frac{1}{x_T} \!
\right) \!\!
\left[ \!
1 \!\pm\!
2 N(0) \!
\frac{\mbox{arcsinh} \!
\left( \!
\frac{1}{x_T} \!
\right)
}
{\sum _p \! A_p}
\frac{\Delta _T \!\cdot\! \Delta _T \!
\sum _p \! B_p
}
{1 \!\pm\! [\det z_{12}]_T ^{\frac{1}{2}}}
\!\cdot\!
[\det z_{12}]_T ^{\frac{1}{2}}\!
\right]
}
\nonumber \\
\nonumber \\
&\!\!\!\!\!\!\!\!\!\!\!\!\!
\!\!\times\!\!
\left[ \!
1
\!\!+\!\!
\left\{ \!
{\displaystyle
\!\mp\! 1
\!\mp\!
\frac{\Delta ^2 _T \! \sum _p \! C_p}
{\sum _p \! A_p}
\!\!\pm\!\! 4 \! N(0) \!
\frac{\mbox{arcsinh} \!
\left( \!
\frac{1}{x_T} \!
\right)
}
{\sum _p \! A_p}
\frac{\Delta_{\!T} \!\cdot\! \Delta _T \!
\sum _p \! B_p
}
{1 \!\pm\! [\det z_{12}]_T ^{\frac{1}{2}}}
} \!\!
\right\}
\! \cdot \!
[\det z_{12}]_T ^{\frac{1}{2}} \!
\right] ^{\!-1}\!\!,
\label{GapequationwithABandC}
\eeqa
which reduces to the Res-HB gap equation
(4.10) in I 
as $T \!\!\rightarrow\!\! 0$.
Using a variable $\varepsilon \!=\! \xi \Delta _T$ 
instead of $\varepsilon$,
the summations
$\sum _p \! A_p,\sum _p \! B_p$ and $\sum _p \! C_p$
near $T \!=\! 0$
can be computed to be\\[-16pt]
\beqa
\!\!\!\!\!\!
\left.
\begin{array}{l}
{\displaystyle
\frac{\sum _p \! A_p}{2 N(0)}
\!=\!\!
\int_{0}^{ \frac{1}{x_T}} \!\!\! d\xi
\frac
{\xi ^2}
{(\xi ^2 \!+\! 1) ^{\frac{3}{2}}}
\!=\!
\mbox{arcsinh} \!
\left( \!
\frac{1}{x_T} \!
\right)
\!-\! \frac{1}{\sqrt{1 \!+\! x ^2 _T}}
} , \\
\\[-8pt]
{\displaystyle
\frac{\Delta_T \sum _p B_p}{N(0)}
\!=\!
2 \arctan \!
\left( \!
\frac{1}{x_T} \!
\right) 
} , ~
{\displaystyle
\frac{\Delta_T ^2 \sum _p C_p}{2 N(0)}
\!=\!
\frac{1}{\sqrt{1 \!+\! x ^2 _T}}
}, 
\end{array}
\right\}
\label{TdependenceofABandC}
\eeqa\\[-12pt]
Substituting (\ref{TdependenceofABandC}) into
(\ref{GapequationwithABandC}) and near $T \!=\! 0$
approximating as\\[-16pt]
\beqa
\!\!\!\!
\left.
\begin{array}{c}
{\displaystyle
\arctan \!
\left( \!
\frac{1}{x_T} \!
\right)
\!\simeq\!
\frac{\pi }{2} \!-\! x_T,~~
\frac{1}{\sqrt{1 \!+\! x ^2 _T}}
\!\simeq\!
1 \!-\! x_0 x_T,~~
(0 \!<\! x_0 \!\ll\! 1)
}\\
\\[-12pt]
{\displaystyle
\mbox{arcsinh} \!
\left( \!
\frac{1}{x_T} \!
\right)
\!\simeq\!
\mbox{arcsinh} \!
\left( \!
\frac{1}{x_0} \!
\right)
\!-\!
\frac{1}{x_0}
(x_T \!-\! x_0),~~
[\det z_{12}]_T ^{\frac{1}{2}}
\!\simeq\!
[\det z_{12}]^{\frac{1}{2}}_{T=0} 
} ,
\end{array} \!
\right\}
\label{Approxmations}
\eeqa\\[-10pt]
then, near $T=0$ we have the gaps for Case I
(4.5) in I as\\[-18pt]
\beqa
\begin{array}{l}
\Delta _T ^{\mbox{\scriptsize I}}
\!=\!
\Delta_0 \!
\left\{ \!
{\displaystyle
1
\!-\!
\frac{1 \!-\! [\det z_{12}]^{\frac{1}{2}}_{T=0}}{N(0) V}
} \!
\right\} \!
\left\{ \!
1 \!+\!
\mbox{arcsinh} \!
\left( \!
{\displaystyle \frac{\hbar \omega_D}{\Delta_0}} \!
\right) \!
\right\} \\
\\[-12pt]
\!\times\!
\left[ \!
{\displaystyle
1
\!-\!
\frac{1 \!-\! [\det z_{12}]^{\frac{1}{2}}_{T=0}}{N(0) V}
} 
\!+\!
\pi N(0) \Delta_0
{\displaystyle
\frac{[\det z_{12}]^{\frac{1}{2}}_{T=0}}
{1 \!+\! [\det z_{12}]^{\frac{1}{2}}_{T=0}}
} \!
\left\{ \!
1 \!+\!
\mbox{arcsinh} \!
\left( \!
{\displaystyle \frac{\hbar \omega_D}{\Delta_0}} \!
\right) \!
\right\} \!
\right]^{\!-1} ,
\end{array}
\label{DeltapnearT0}
\eeqa\\[-8pt]
and
Case II
(4.6) in I
with the aid of
$
[\det z_{12}]^{1 / 2}_{T=0}
\!\simeq\!
1 \!-\! 2 \pi \! N(0) \hbar \omega_D x_0
(0 \!<\! x_0 \!\ll\! 1)
$
easily derived from Taylor expansion of
(\ref{detz}),
as\\[-16pt]
\beqa
\!\!\!\!
\begin{array}{c}
\Delta _T ^{\mbox{\scriptsize II}}
\!=\!
\left\{ \!
{\displaystyle
\hbar \omega_D
\!-\!
\frac{\Delta_0}
{\pi \!\!-\!\! 1 \!\!-\!\! \frac{\pi }{2} \!\cdot\! N(0) V
\!\!-\!\!
( \! 1 \!\!-\!\! N(0) V \! )
\!\cdot\!
[\det z_{12}]^{- \frac{1}{2}}_{T=0} \!
} \!
\! + \!
\Delta_0
\mbox{arcsinh}  \!
\left( \!\!
\frac{\hbar \omega_D}{\Delta_0} \!\!
\right) 
} \!\!
\right\} \! .
\end{array}
\label{DeltamnearT0}
\eeqa\\[-16pt]

In the opposite limit
$
T
\!\!\rightarrow\!\!
T_c ^{\mbox{\scriptsize I}}
$
($T_c$ for Case I)
the gap becomes very small,
$
[\det z_{12}]_T ^{1 / 2}
\!\!\rightarrow\!\! 1
$,
then
$
{\cal F}_{\! \Delta _T} ^{\uparrow }
\!\!\rightarrow\!\!
-
\Delta
N(0) V 
\mbox{arcsinh} \!
\left(
\hbar \omega_D / \Delta
\right) \!
/ 4
$
and
$
({\cal F}^{\uparrow }_{\! + \varepsilon_p T}
\!\!+\!\!
{\cal F}^{\uparrow }_{\! - \varepsilon_p T})
/ 2
\!\!\rightarrow\!\!
\varepsilon _p
/ 4
$
if we use the last relation
in 
(5.9)
and the second one
in 
(5.16) in I.
We have an approximate QP energy
$
\widetilde{\epsilon }_p ^{\uparrow \mbox{\scriptsize I}}
\!\simeq\!
\sqrt{
\varepsilon _p ^2 
\!+\! 
\left\{ \!
-\Delta 
N(0) V
\mbox{arcsinh} \!
\left( \!
\hbar \omega_D / \Delta
\right) \!
\right\} ^2
} \!
/ 4
$.
This is because
the two HB WFs
have different correlation structures
${\psi }_{2} \!=\! \pi$ and ${\psi }_{1} \!=\! 0$.
In such a case,
returning to the original form of the BCS gap equation but with
the modified QP energy
$\widetilde{\epsilon }_p ^{\uparrow \mbox{\scriptsize I}}$,
the thermal gap equation is expressed as 
$
1 
\!=\!
V / 2
\sum_p
\left( 1 \!-\! 2 \widetilde{w}_{p} ^{\uparrow } \right) \!
/ \widetilde{\epsilon }_p ^{\uparrow \mbox{\scriptsize I}}
$
and leads to the integral form\\[-12pt]
\beqa
\begin{array}{c}
1
{\displaystyle
\!=\! 
\frac{V}{2} \!
}
\sum_p 
{\displaystyle
\frac{4}{\varepsilon_p}
\tanh \!
\left( \!
\frac{\varepsilon_p}{8k_B T ^{\mbox{\scriptsize I}}} \!
\right)
\!=\!
4 N(0) V \!\! \int _0 ^{\hbar \omega_D} \!\!\! d \varepsilon
\frac{1}{\varepsilon }
\tanh \!
\left( \!
\frac{\varepsilon }{8k_B T ^{\mbox{\scriptsize I}}} \!
\right) .
}
\end{array}
\label{gapequation3}
\eeqa
Introduce a dimensionless variable
$
y_T ^{\mbox{\scriptsize I}}
\!\!\equiv\!\!
\varepsilon  / 8k_B T ^{\mbox{\scriptsize I}}
$
and its upper-value 
$
y ^{\mbox{\scriptsize I}}_{T_c ^{\mbox{\scriptsize I}}}
\!\!\equiv\!\!
\hbar \omega_D / 8k_B T_c ^{\mbox{\scriptsize I}}
$.
Integrating R.H.S. of
(\ref{gapequation3})
 by parts, 
it is approximated as follows:
\beqa
\begin{array}{ll}
\!\!\!\!
{\displaystyle \frac{1}{4 N(0) V}}
&\!\!\simeq
\ln
y ^{\mbox{\scriptsize I}}_{T_c ^{\mbox{\scriptsize I}}}
\! - \!
{\displaystyle
\! \int _0 ^{\infty } \! \!\! dy
\ln y~\mbox{sech} ^2 y
} 
\! = \!
\ln
y ^{\mbox{\scriptsize I}}_{T_c ^{\mbox{\scriptsize I}}}
\! + \!
\ln \!
\left( \!
{\displaystyle
\frac{4 e^{\gamma }}{\pi } \!
}
\right) \\
\\[-6pt]
&\!\! = \!
\ln \!
\left( \!
{\displaystyle
\frac{e^{\gamma }}{2 \pi }
\frac{\hbar \omega_D}{k_B T_c ^{\mbox{\scriptsize I}}} \!
}
\right)
\!\equiv\!
\ln \!
\left( \!
{\displaystyle
\frac{\theta _D}{\widetilde{T}_c ^{\mbox{\scriptsize I}}} \!
}
\right), 
\left( \!
\theta _D
\!\equiv\!
{\displaystyle
\frac{\hbar \omega_D}{k_B}
} ~ \mbox{:Debye temperature} \!
\right),
\end{array}
\label{gapequation4}
\eeqa
where we have used the formula in the textbook
\cite{GraRyz.63}.
Number $\gamma$ is
the Euler's constant 
$(\gamma \!\simeq\! 0.5772)$ 
and
$e^{\gamma } \!\simeq\! 1.781$.
Finally a small rearrangement yields\\[-20pt]
\beqa
\!\!
T_c ^{\mbox{\scriptsize I}}
\!=\!
 0.283 \theta _D
e ^{
- \frac{1}{4 N(0) V}
} ,
\label{criticalTemp}
\eeqa\\[-20pt]
which should be compared with
the Eliashberg's formula
\cite{Eliashberg.61}
and
the usual HB's one 
for $T_c$\\[-20pt]
\beqa
T_c\!=\!
 1.130 \theta _D
e ^{
- \frac{1}{N(0) V}
} .
\label{HBcriticalTemp}
\eeqa\\[-20pt]
The new formula
(\ref{criticalTemp})
gives a {\it high} critical temperature,
e.g.,
$T_c ^{\mbox{\scriptsize I}} \!=\! 72.87 $K
for $N(0)V \! = \! 0.25$ and $\theta _D \!=\! 700 $K.
This $T_c ^{\mbox{\scriptsize I}}$ 
is in contrast to $T_c$
obtained by the usual HB's
(\ref{HBcriticalTemp}),
i.e., $T_c \!=\! 14.49 $K for the same values
of $N(0)V$ and $\theta _D$.

We are now in a stage to discuss
the behaviour of the gap near $\!T_{\!c}$.$\!$
In the above the modified QP energy
$\!
\widetilde{\epsilon }
\!=\!\!
\sqrt{\varepsilon ^2
\!+\!
\left\{ \!
-\Delta \!
N \!(0) V \!
\mbox{arcsinh} \!
\left(
\hbar \omega_D / \Delta
\right) \!
\right\}^2} \!
/ 4
$
plays a crucial role to boost the $T_{\!c}$
in (\ref{criticalTemp})
comparing with 
the numerical result in
(\ref{HBcriticalTemp}).
Notice the exsistence of the numerical factor
$1 / 4$.
Now let us consider
$\Delta^{\mbox{\scriptsize I}} _T$
near $T_{\!c} ^{\!\mbox{\scriptsize I}}$.
Using this form of QP energy,
the gap equation is roughly rewritten as\\[-12pt]
\beqa
\!\!\!\!\!\!\!\!
\begin{array}{c}
{\displaystyle
\frac{1}{4 N(0) V}
\!=\!
\frac{1}{4}
\int _0 ^{\hbar \omega_D} \! d \varepsilon
\frac{1}{\widetilde{\epsilon }}
\tanh \!
\left( \!\!
\frac{\widetilde{\epsilon }}{2k_B T} \!\!
\right) 
\!\simeq\! 
\int _0 ^{\hbar \omega_D} \! d \varepsilon
\frac{1}{\varepsilon }
\tanh \!
\left( \!\!
\frac{\varepsilon }{8k_B T} \!\!
\right)
}\\
\\[-6pt] 
{\displaystyle
- \!
\left\{ \!
\Delta _{\!T} \!
N(0) V \!
\! \cdot \! \mbox{arcsinh} \!
\left( \!\!
{\displaystyle \frac{\hbar \omega_D}{\Delta _{\!T}}} \!\!
\right) \!\!
\right\} ^{\!2} \!
\!\!
\int _0 ^{\hbar \omega_D} \!\!\!\!\! d \varepsilon \!
\left\{ \!
\frac{1}{\varepsilon ^3} \!
\tanh \!
\left( \!\!
\frac{\varepsilon }{8k_B T} \!\!
\right)
\!\!-\!\!
\frac{1}{\varepsilon ^2}
\frac{1}{8k_B T}
\mbox{sech} ^{\!2} \!
\left( \!\!
\frac{\varepsilon }{8k_B T} \!\!
\right) \!\!
\right\} \! ,
}
\end{array}
\label{gapequation5}
\eeqa
from which we obtain\\[-16pt]
\beqa
\begin{array}{ll}
& \!\!\!\!\!\!\!\!\!\! 
{\displaystyle
\frac{1}{4 N(0) V}
}
\! = \!
\ln \!
\left( \!
{\displaystyle
\frac{\hbar \omega_D}{k_B \widetilde{T}}
} \!
\right)
\!-\!
{\displaystyle \frac{7}{8 \pi ^2}}
\zeta (3) \!
\left( \!
{\displaystyle
\frac{2 \pi }{e^\gamma }
} \!
\right) ^2 \!
\left( \!
{\displaystyle
\frac{\hbar \omega_D}{k_B \widetilde{T}}
} \!
\right) ^2 \!
\left\{ \!
N(0) V
x_T
\! \cdot \! \mbox{arcsinh} \!
\left( \!
{\displaystyle
\frac{1}{x_T}
} \!
\right) \!
\right\}^2 \\
\\[-6pt]
& \!\!\!\!\!\!\!\!\!\!
\! \simeq \!
\ln \!
\left( \!
{\displaystyle
\frac{\hbar \omega_D}{k_B \widetilde{T}_c ^{\mbox{\scriptsize I}}}
} \!
\right)
\!+\!
{\displaystyle
\frac{\widetilde{T}_c ^{\mbox{\scriptsize I}} \! - \! \widetilde{T}}
{\widetilde{T}_c ^{\mbox{\scriptsize I}}}
}
\!-\!
{\displaystyle \frac{7}{8 \pi ^2}}
\zeta (3) \!
\left( \!
{\displaystyle
\frac{2 \pi }{e^\gamma }
} \!
\right) ^2 \!\!
\left( \!
1 
\! - \!
{\displaystyle
\frac{\widetilde{T}_c ^{\mbox{\scriptsize I}} \! - \! \widetilde{T}}
{\widetilde{T}_c ^{\mbox{\scriptsize I}}}
} \!
\right) ^{-2}\\
\\[-6pt]
&
~~~~~~~~~~~~~~~~~~~~~~~~~~~~~~~~~~~~~~~
\times\!
\left( \!
{\displaystyle
\frac{\hbar \omega_D}{k_B \widetilde{T}_c ^{\mbox{\scriptsize I}}}
} \!
\right) ^2 \!\!
\left\{ \!
N(0) V
x_T
\! \cdot \! \mbox{arcsinh} \!
\left( \!
{\displaystyle
\frac{1}{x_T}
} \!
\right) \!
\right\}
^2 \! ,
\end{array}
\label{gapequation6}
\eeqa\\[-10pt]
where 
$
\hbar \omega_D / k_B \widetilde{T}
\!\equiv\!
e^{\gamma } / 2 \pi
\!\cdot\!
\hbar \omega_D / k_B T
$.
For details see Appendix C.
Using
$
\mbox{arcsinh} \!
\left( \!
1 / x_T \!
\right)
\!\simeq\!
\ln
\left( \!
2 / x_T \!
\right)
\!\simeq\!
\!-\!
\left( x_T \!-\! 2 \right) \! / 2
+ \cdots ~
\left(
2 / x_T
\!>\! 
1 / 2
\right)
$,
(\ref{gapequation4})
and
(\ref{gapequation6}),
we get $\Delta _T^{\mbox{\scriptsize I}}$
near $T_c ^{\mbox{\scriptsize I}}$ as\\[-14pt]
\beqa
{\displaystyle
\Delta _T^{\mbox{\scriptsize I}}
\!\simeq\!
2 \pi \sqrt{\frac{2}{7 \zeta (3)}}
\frac{k_B T_c ^{\mbox{\scriptsize I}}}{N(0) V}
\left( \!
1
\!-\!
\frac{T_c ^{\mbox{\scriptsize I}} - T}
{T_c ^{\mbox{\scriptsize I}}} \!
\right) \!
\sqrt{ \frac{T_c ^{\mbox{\scriptsize I}} - T}
{T_c ^{\mbox{\scriptsize I}}}}
}.
\label{gappnearTc}
\eeqa\\[-12pt]
The $\sqrt{T_c ^{\mbox{\scriptsize I}} \!-\! T}$
dependence of
$\Delta _T^{\mbox{\scriptsize I}}$
is more complicated than the usual one
\cite{AGD.65,Abrikosov.88}.

For Case II,
$
(\!{\cal F}^{\uparrow }_{\!\! + \varepsilon_p T}
\!+\!
{\cal F}^{\uparrow }_{\!\! - \varepsilon_p T}\!)
$,
${\cal F}_{\!\! \Delta _T} ^{\uparrow }$ and
$\widetilde{\epsilon }_p ^{\uparrow }$
become infinite simultaneously in the limit
$\Delta _T \!\!\rightarrow\!\! 0\!~(x_{\!T}
\!\!\rightarrow\!\! 0)$
due to the existence of
$1 \!-\! [\det z_{12}]_T ^{1 / 2}$
in denominator.
Then mathematical handling for such a problem is too difficult and
therefore we can not easily get a formula
for $T_c ^{\mbox{\scriptsize II}}$
in an analytical way as we did in Case I.
Denote $T_c$ for Case II as
$T_c ^{\mbox{\scriptsize II}}$.
At $T \!\!\simeq\!\! T_c ^{\mbox{\scriptsize II}}$,
$\Delta _T^{\mbox{\scriptsize II}}$ almost vanishes
and 
$
1 \!-\! [\det z_{12}]_T ^{1 / 2}
\!\!\rightarrow\!\!
2 \pi N(0) \hbar \omega_{\!D} x_T
$.
Using
${\cal F}_{\! \Delta _T} ^{\uparrow }$ in
(\ref{ResFepandFdp2}),
we reach to the following asymptotic forms:
$
{\cal F}_{\! \Delta _T} ^{\uparrow }
\!\!\rightarrow\!\!
-
\left( 4 \pi N(0) \right)^{-1} \!\!
N(0) V 
\mbox{arcsinh} \!
\left(
\hbar \omega_{\!D} / \Delta _{\!T}
\right)
$
and
$
({\cal F}^{\uparrow }_{\! + \varepsilon_p T}
\!+\!
{\cal F}^{\uparrow }_{\! - \varepsilon_p T})
/ 2
\rightarrow
\left( 4 \pi N(0) \right)^{-1} \!
\varepsilon _p / \Delta _T
$.
The QP energy
$
\widetilde{\epsilon }_p ^{\uparrow }
\left(
\!=\!
\sqrt{
\left( \!
{\cal F}^{\uparrow }_{\! + \varepsilon_p T}
\!+\!
{\cal F}^{\uparrow }_{\! - \varepsilon_p T} \!
\right)^2 \!\!
/ 4
\!+\!
{\cal F}_{\! \Delta _T} ^{\uparrow 2}
} \!
\right)
$
is approximately calculated to be
$
\widetilde{\varepsilon }_p ^{\uparrow \mbox{\scriptsize II}}
\!\simeq\!
\left( 4 \pi N(0) \right)^{-1} \!
\varepsilon _p / \Delta _T ~
\left( 0 \! < \! \Delta _T \! \ll 1 \right)
$.
Here we discard the contribution from
${\cal F}_{\! \Delta _T} ^{\uparrow }$
comparing with the one from
$
({\cal F}^{\uparrow }_{\! + \varepsilon_p T}
\!+\!
{\cal F}^{\uparrow }_{\! - \varepsilon_p T})
/ 2$.
As was done previously,
returning again to the original form of the BCS gap equation but with
another modified QP energy
$\widetilde{\epsilon }_p ^{\uparrow \mbox{\scriptsize II}}$,
the thermal gap equation is obtained as 
$
1 
\!=\!
V / 2
\sum_p
\left( 1 \!-\! 2 \widetilde{w}_{p} ^{\uparrow } \right) \! /
\widetilde{\epsilon }_p ^{\uparrow \mbox{\scriptsize II}}
$
which also leads to the integral form\\[-10pt]
\beqa
\begin{array}{cc}
&1
\!=\! 
{\displaystyle \frac{V}{2}} \!
\sum_p 
{\displaystyle
\frac{4 \pi N(0) \Delta _T ^{\mbox{\scriptsize II}}}{\varepsilon _p}
}
\tanh \!
\left( \!
{\displaystyle
\frac{\varepsilon _p}
{2k_B T ^{\mbox{\scriptsize II}} \!\cdot\! 4 \pi N(0)
\Delta _T ^{\mbox{\scriptsize II}}}
} \!
\right) \\
\\[-4pt]
&\!=\!
{\displaystyle
N(0) V \!\! \int _0 ^{\hbar \omega_D} \!\!\! d \varepsilon
\frac{4 \pi N(0) \Delta _T ^{\mbox{\scriptsize II}}}{\varepsilon }
}
\tanh \!
\left( \!
{\displaystyle
\frac{\varepsilon }
{2k_B T ^{\mbox{\scriptsize II}} \!\cdot\! 4 \pi N(0)
\Delta _T ^{\mbox{\scriptsize II}}}
} \!
\right) .
\end{array}
\label{gapequation7}
\eeqa\\[-10pt]
Introduce a dimensionless variable
$
y_T ^{\mbox{\scriptsize II}}
\!\!\equiv\!\!
\varepsilon / \!
\left( 
2 k_B T \!\cdot\! 4 \pi N(0)
\Delta _T ^{\mbox{\scriptsize II}}
\right)
$
and its upper-value 
$
y ^{\mbox{\scriptsize II}}_{T ^{\mbox{\scriptsize II}}}
\!\!\equiv\!\!
\hbar \omega_D / \!
\left(
2 k_B T ^{\mbox{\scriptsize II}} \!\cdot\! 4 \pi N(0) \Delta _T ^{\mbox{\scriptsize II}}
\right)
$.
Integrating the last equation in
(\ref{gapequation7})
by parts, 
it is approximately calculated as\\[-16pt]
\beqa
\begin{array}{ll}
{\displaystyle \frac{1}{4 N(0) V}}
{\displaystyle \frac{1}{\pi N(0) \Delta _T ^{\mbox{\scriptsize II}}}}
&\!\!\!\!\simeq\!
\ln
y ^{\mbox{\scriptsize II}}_{T ^{\mbox{\scriptsize II}}}
\! + \!
\ln \!
\left( \!
{\displaystyle
\frac{4 e^{\gamma }}{\pi } \!
}
\right) \\
\\[-6pt]
&\!\! = \!
\ln \!
\left( \!
{\displaystyle
\frac{1}{\pi N(0) \Delta _T ^{\mbox{\scriptsize II}}}
\frac{e^{\gamma }}{2 \pi }
\frac{\hbar \omega_D}{k_B T ^{\mbox{\scriptsize II}}} \!
}
\right)
\!=\!
\ln \!
\left( \!
{\displaystyle
\frac{1}{\pi N(0) \Delta _T ^{\mbox{\scriptsize II}}}
\frac{\theta _D}{\widetilde{T} ^{\mbox{\scriptsize II}}} \!
}
\right) ,
\end{array}
\label{gapequation8}
\eeqa
which reads\\[-12pt]
\beqa
\!\!\!\!
\Delta _T ^{\mbox{\scriptsize II}}
\!=\!
{\displaystyle
\frac{\theta _D}{\pi \! N \! (0) \widetilde{T} ^{\mbox{\scriptsize II}}} \!
}
\exp \!
\left\{ \!\!
{\displaystyle
-\frac{1}{4 \! N \! (0) \! V \! \pi \! N(0) \!
\Delta _T ^{\mbox{\scriptsize II}}}
} \!\!
\right\} 
\!\approx\!
{\displaystyle
\frac{\theta _D}{\pi \! N \! (0) \widetilde{T} ^{\mbox{\scriptsize II}}} \!
}
\left(\!\!
1
\!-\!
{\displaystyle 
\frac{1}{4 \! N \! (0) \! V \! \pi \! N(0) \!
\Delta _T ^{\mbox{\scriptsize II}}}
} \!\!
\right) ,
\label{eqdelT1}
\eeqa
from which we obtain an equation to determine
$\Delta _T ^{\mbox{\scriptsize II}}$ very near $T_c$ as\\[-12pt]
\beqa
\Delta _T  ^{\mbox{\scriptsize II}2}
\!-\!
{\displaystyle
\frac{1}{\pi N(0)}
\frac{\theta _D}{\widetilde{T} ^{\mbox{\scriptsize II}}}
}
\Delta _T ^{\mbox{\scriptsize II}}
\!+\!
{\displaystyle
\frac{1}{\pi N(0)}
\frac{\theta _D}{\widetilde{T} ^{\mbox{\scriptsize II}}}
}
{\displaystyle 
\frac{1}{4 N (0) V \pi N(0)}
}
\!=\!
0 .
\label{eqdelT2}
\eeqa
Then we have a solution for
$\Delta _T ^{\mbox{\scriptsize II}}$
as\\[-16pt]
\beqa
\Delta _T ^{\mbox{\scriptsize II}}
\!=\!
{\displaystyle
\frac{1}{\pi N(0)}
\frac{\theta _D}{\widetilde{T} ^{\mbox{\scriptsize II}}}
}
\!-\!
{\displaystyle 
\frac{1}{4 N (0) V \pi N(0)}
} ,
\label{soldelT}
\eeqa
in which at
$
\Delta _T ^{\mbox{\scriptsize II}}
\!=\!
T_c ^{\mbox{\scriptsize II}}
$,
the
$\Delta _T ^{\mbox{\scriptsize II}}$
vanishes.
Then finally we can determine the critical temperature
$\Delta _{T _c} ^{\mbox{\scriptsize II}}$
for Case II as\\[-16pt]
\beqa
T_c ^{\mbox{\scriptsize II}}
\!=\!
{\displaystyle 
\frac{2 e^\gamma }{\pi }
}
\theta _D
N (0) V 
\!=\!
1.334
\theta _D
N (0) V .
\label{TcforII}
\eeqa
The simple formula
(\ref{TcforII})
gives a {\em high} critical temperature, e.g.,
$T_c ^{\mbox{\scriptsize II}}
\!=\!
$
198K for
$N (0) V 
\!=\!
$
0.25
and
$\theta _D
\!=\!
$
700K.
Finally
$\Delta _T ^{\mbox{\scriptsize II}}$
near
$T_c ^{\mbox{\scriptsize II}}$
can be approximately obtained as\\[-16pt]
\beqa
\Delta _T ^{\mbox{\scriptsize II}}
\!\approx\!
-
{\displaystyle
\frac{e^\gamma }{2 \pi }
\frac{1}{\pi N(0)}
\frac{\theta _D}{T_c ^{\mbox{\scriptsize II}}}
\frac{T \!-\! T_c ^{\mbox{\scriptsize II}}}
{T_c ^{\mbox{\scriptsize II}}}
} ,
\eeqa
which is linearly dependent on
$T \!\!-\!\! T_c ^{\mbox{\scriptsize II}}$.
It is very interesting that we could find
such a dependence of $\Delta _T^{\!\mbox{\scriptsize II}}$,
comparing with the usual dependence
$\!\sqrt{T \!\!-\!\! T_c ^{\mbox{\scriptsize II}}}\!$
of $\Delta _T^{\!\mbox{\scriptsize II}}$.

In intermediate temperature region
the modified QP energy $\widetilde{\varepsilon }$
is approximated as
$
\widetilde{\varepsilon } ^{(\pm)}
\!=\!
\sqrt{\varepsilon ^2 \!+\! \widetilde{\Delta }_T ^2}
\left\{ \!
2 \left( 1 \!\pm\! [\det z_{12}]_T ^{\frac{1}{2}} \right) \!
\right\}^{-1}
$.
When
$\varepsilon \!\gg\! \widetilde{\Delta }_T$,
each term
$\sum _p \! A_p,~\sum _p \! B_p$ and
$\sum _p \! C_p$
in
(\ref{DefinitionsofABandC})
is approximately computed as\\[-10pt]
\beqa
\!\!\!\!\!\!\!\!
\left.
\begin{array}{c}
{\displaystyle
\frac{\sum _p \! A_p}{2 N(0)}
\!\! = \!\!
\int _{0} ^{\hbar \omega_D} \!\!\!\!\! d \varepsilon
\frac
{\varepsilon ^2}
{
\left( \!
\varepsilon  ^2
\!\!+\!\!
\widetilde{\Delta }_T ^2 \!
\right)^{\frac{3}{2}}
}
\tanh \!
\left( \!\!
\frac{\widetilde{\varepsilon } ^{(\pm)}}{2k_B T} \!\!
\right)
}
{\displaystyle
= \!
\ln \!
\left( \!
\frac{4 e ^\gamma }{\pi }
y_T ^{(\pm)} \!
\right)
\! - \!
\frac{21}{2 \pi ^2}
\zeta (3) \!
\left( \!
y_T ^{(\pm)}
\widetilde{x}_T \!
\right) ^{\!2}
},\\
\\[-6pt]
{\displaystyle
\frac{\widetilde{\Delta }_T \! \sum _p \! B_p}{N(0)}
}
\!\! = \! 0,
{\displaystyle
\frac{\widetilde{\Delta }_T ^2 \! \sum _p \! C_p}{2 N(0)}
\!\! = \!\!
\int _{0} ^{\hbar \omega_D} \!\!\!\!\! d \varepsilon
\frac
{1}
{
\left( \!
\varepsilon ^2
\!\!+\!\!
\widetilde{\Delta }_T ^2 \!
\right)^{\!\frac{3}{2}}
} \!
\tanh \!
\left( \!\!
\frac{\widetilde{\varepsilon }^{(\pm)}}{2k_B T} \!\!
\right)
}
{\displaystyle
= \!\!
\frac{7}{\pi ^2}
\zeta (3) \!
\left( \!
y_T ^{\mbox{\scriptsize II}}
x_T \!
\right) ^{\!2}
} \! ,
\end{array} \!\!
\right\}
\label{DefinitionsofABandC2}
\eeqa
whose details are presented in Appendix C.
Taking only a leading term,
$A_p$ and $C_p$ terms in
(\ref{DefinitionsofABandC2})
are approximated to be\\[-16pt]
\beqa
\!\!\!\!\!\!\!\!
\left.
\begin{array}{l}
{\displaystyle
\frac{\sum _p \! A_p}{2 N(0)}
\!\simeq\!
\ln \!
\left( \!
\frac{4 e ^\gamma }{\pi }
y_T ^{(\pm)} \!
\right)
\!-\!
\frac{21 \zeta (3)}{2 \pi ^2} \!
\left( \!
y_T ^{(\pm)}
\widetilde{x}_T \!
\right) ^{\!2}
\!\!=\!
\ln \!
\left( \!\!
\frac{ e ^\gamma }{\pi }
\frac{1}{1 \!\!\pm\!\! [\det z_{12}]_T ^{\frac{1}{2}}}
\frac{\theta_D}{T} \!\!
\right)
\!-\!
\alpha _T ^{(\pm)}
} ,\\
\\[-10pt]
{\displaystyle
\frac{\widetilde{\Delta }_T ^2 \! \sum _p \! C_p}{2 N(0)}
\!\simeq\!
\frac{7 \zeta (3)}{\pi ^2} \!
\left( \!
y_T ^{(\pm)}
\widetilde{x}_T \!
\right) ^2 \!
\!=\!
\frac{2}{3}
\alpha_T ^{(\pm)} \!
} , \\
\\[-10pt]
{\displaystyle
\alpha_T ^{(\pm)}
\!\equiv\!
\frac{21 \zeta (3)}{2 \pi ^2} \!
\left( \!
\frac{ e ^\gamma }{\pi }
\frac{1}{1 \!\!\pm\!\! [\det z_{12}]_T ^{\frac{1}{2}}} \!\!
\right) ^{\!2} \!\!
\left( \!
\frac{\widetilde{\Delta }_T}{k_B T} \!
\right) ^{\!2} \!
} \! , ~
\widetilde{\Delta }_T
\!\equiv\!
\Delta _T N(0) V
\mbox{arcsinh} \!
\left( \!\!
{\displaystyle \frac{\hbar \omega_D}{\Delta_T}} \!\!
\right) .
\end{array} \!\!
\right\}
\label{DefinitionsofABandC3}
\eeqa\\[-8pt]
Substituting these results into
(\ref{GapequationwithABandC}),
we have\\[-18pt]
\beqa
\!\!\!\!\!\!\!\!
\begin{array}{l}
\left( \!\!
{\displaystyle
\frac{ e ^\gamma }{\pi }
\frac{1}{1 \!\!\pm\!\! [\det z_{12}]_T ^{\frac{1}{2}}}
} \!\!
\right) ^{\!\!2} \!\!
x_T ^2 \!
\left\{ \!\!
N(0) V \! \mbox{arcsinh} \!
\left( \!\!
{\displaystyle
\frac{1}{x_T}
} \!\!
\right) \!\!
\right\}^{\!\!2} \!\!
\left\{ \!\!
N(0) V \! \mbox{arcsinh} \!
\left( \!\!
{\displaystyle
\frac{1}{x_T}
} \!\!
\right) \!
\!-\!\!
\left( \!\!
1 \!\!-\!\! [\det z_{12}]_T ^{\frac{1}{2}} \!\!
\right) \!\!
\right\} \!
 \\
\\[-8pt]
-
{\displaystyle
\frac{2 \pi ^2}{21 \zeta (3)}
} \!
\ln \!
\left( \!
{\displaystyle
\frac{ e ^\gamma }{\pi }
\frac{1}{1 \!\!\pm\!\! [\det z_{12}]_T ^{\frac{1}{2}}}
\frac{\theta_D}{T}
} \!
\right) \!\!
\left( \!
{\displaystyle
\frac{T}{\theta_D}
} \!
\right) ^{\!\!2} \!\!
N(0) V \! \mbox{arcsinh} \!
\left( \!\!
{\displaystyle
\frac{1}{x_T}
} \!\!
\right) \\
\\[-8pt]
+
{\displaystyle
\frac{2 \pi ^2}{21 \zeta (3)}
} \!\!
\left( \!
1 \!-\! [\det z_{12}]_T ^{\frac{1}{2}} \!
\right) \!
\ln \!
\left( \!
{\displaystyle
\frac{ e ^\gamma }{\pi }
\frac{1}{1 \!\!\pm\!\! [\det z_{12}]_T ^{\frac{1}{2}}}
\frac{\theta_D}{T}
} \!
\right) \!\!
\left( \!
{\displaystyle
\frac{T}{\theta_D}
}
\right) ^{\!\!2}
\\
\\[-8pt]
\!\mp\!
{\displaystyle \frac{2}{3}}
\left( \!\!
{\displaystyle
\frac{ e ^\gamma }{\pi }
\frac{1}{1 \!\!\pm\!\! [\det z_{12}]_T ^{\frac{1}{2}}}
} \!\!
\right) ^{\!\!2} \!\!
x_T ^2
[\det z_{12}]_T ^{\frac{1}{2}}
=
0 ,
\end{array}
\label{determinationofcriticaltemperature}
\eeqa\\[-8pt]
to be solved analytically for a given $T$, 
which is rewritten as
\beqa
\!\!\!\!
\begin{array}{l}
\left[ \!
\left( \!\!
{\displaystyle
\frac{ e ^\gamma }{\pi }
\frac{1}{1 \!\!\pm\!\! [\det z_{12}]_T ^{\frac{1}{2}}}
} \!\!
\right) ^{\!2} \!
x_{T} ^2 \!
\left\{
N(0) V \! \mbox{arcsinh} \!
\left( \!
{\displaystyle
\frac{1}{x_{T}}
} \!
\right)
\right\}^{2} 
\right. \\
\\[-8pt]
\!\!
\left.
\!-
{\displaystyle
\frac{2 \pi ^2}{21 \zeta (3)}
} \!
\ln \!
\left( \!\!
{\displaystyle
\frac{ e ^\gamma }{\pi }
\frac{1}{1 \!\!\pm\!\! [\det z_{12}]_T ^{\frac{1}{2}}}
\frac{\theta_D}{T}
} \!\!
\right) \!\!
\left( \!
{\displaystyle
\frac{T}{\theta_D}
} \!
\right) ^{2}
\right] \!\!
\left\{ \!
N(0) V \! \mbox{arcsinh} \!
\left( \!
{\displaystyle
\frac{1}{x_{T}}
} \!
\right) \!
\!-\!\!
\left( \!
1 \!\!-\!\! [\det z_{12}]_T ^{\frac{1}{2}} \!
\right) \!
\right\} \\
\\[-8pt]
\!\!
=
\!\pm\!
{\displaystyle \frac{2}{3}}
\left( \!\!
{\displaystyle
\frac{ e ^\gamma }{\pi }
\frac{1}{1 \!\!\pm\!\! [\det z_{12}]_T ^{\frac{1}{2}}}
} \!\!
\right) ^{\!2} \!
x_T ^2
[\det z_{12}]_T ^{\frac{1}{2}}
\approx
0 ,~(0 \!<\! x_T \!\ll\! 1) \\
\end{array}
\label{determinationofcriticaltemperature2}
\eeqa\\[-12pt]
from which we obtain an equation to determine
$\Delta_T$
for a given $T$ as\\[-6pt]
\beqa
\!\!
\begin{array}{l}
{\displaystyle
N(0) V 
\frac{ e ^\gamma }{\pi }
\frac{1}{1 \!\pm\! [\det z_{12}]_T ^{\frac{1}{2}}}
}
x_{T}
\mbox{arcsinh}
\left(
{\displaystyle
\frac{1}{x_{T}}
}
\right) \\
\\[-8pt]
\!=\!
-
\sqrt{ \!
{\displaystyle
\frac{2 \pi ^2}{21 \zeta (3)}
} \!
}
\sqrt{
\ln \!
\left( \!
{\displaystyle
\frac{ e ^\gamma }{\pi }
\frac{1}{1 \!\pm\! [\det z_{12}]_T ^{\frac{1}{2}}}
} \!
\right)
\!-\!
\ln
\left(
{\displaystyle
\frac{T}{\theta_D}
}
\right) \!
}
\cdot
\left(
{\displaystyle
\frac{T}{\theta_D}
}
\right) .
\end{array}
\label{determinationofcriticaltemperature3}
\eeqa
This is classified into the following two cases:

Case I:
$[\det z_{12}]_T ^{\frac{1}{2}}
\approx
0.3
$
\beqa
\begin{array}{r}
0.436
N(0) V
x_{T}
\mbox{arcsinh}
\left(
{\displaystyle
\frac{1}{x_{T}}
}
\right)
\!=\!
-
\sqrt{0.782}
\sqrt{ \!
\ln (0.436)
-
\ln
\left(
{\displaystyle
\frac{T}{\theta_D}
}
\right) \!
}
\cdot
\left(
{\displaystyle
\frac{T}{\theta_D}
}
\right) ,
\end{array}
\label{CaseI}
\eeqa
from which,
using the approximate relation
$
\mbox{arcsinh} \!
\left( \!
1 / x_{T} \!
\right)
\!\approx\!
\left( x_{T} \!-\! 2 \right) \! / 2
$,
finally we have a solution for
$x_T~(N(0) V \!=\! 0.25~\mbox{and}~\theta_{D} \!=\! 700\mbox{K})$
as
\beqa
\!\!\!\!\!\!
\begin{array}{r}
x_{T}
\!=\!
1
\!-\!
{\displaystyle
\frac{1}{\sqrt{0.436 \!\times\! 0.25}}
} \!
\sqrt{ \!
0.436 \!\times\! 0.25
\!-\!
2 \! \sqrt{\! 0.782} \!
\sqrt{ \!
\ln (0.436)
\!-\!
\ln \!
\left( \!
{\displaystyle
\frac{T}{700}
} \!
\right)
}
\!\cdot\!
\left( \!
{\displaystyle
\frac{T}{700}
} \!
\right)
} ,
\end{array}
\label{solCaseI}
\eeqa

Case II:
$ 1 - [\det z_{12}]_T ^{\frac{1}{2}}
\approx
2 \pi N (0) \hbar \omega_{D} x_T~\mbox{and}~
\ln
\left( \!
{\displaystyle
\frac{1}{x_{T}}
} \!
\right)
\approx
\mbox{arcsinh} \!
\left( \!
{\displaystyle
\frac{1}{x_{T}}
} \!
\right)
$
\beqa
\begin{array}{r}
(0.283)^2 \!
\left\{ \! N(0) V \! \right\}^2 \!
\mbox{arcsinh}^2 \!
\left( \!
{\displaystyle
\frac{1}{x_{T}}
} \!
\right)
\!-\!
0.782
\left\{ \! \pi \! N(0) \hbar \omega_D \! \right\}^2 \!
\left( \!
{\displaystyle
\frac{T}{\theta_D}
} \!
\right)^2 \!\!
\mbox{arcsinh} \!
\left( \!
{\displaystyle
\frac{1}{x_{T}}
} \!
\right) \\
\\[-12pt]
-
0.782 \!
\left\{ \! \pi \! N(0) \hbar \omega_D \! \right\}^2
\left[ \!\!\! {}^{{}^{{}^{{}^{{}^{.}}}}}
\ln (0.283)
\!-\!
\ln
\left\{ \! 2 \pi \! N(0) \hbar \omega_D \! \right\}
\!-\!
\ln
\left( \!
{\displaystyle
\frac{T}{\theta_D} \!
}
\right)
\right]
\!\cdot\!
\left( \!
{\displaystyle
\frac{T}{\theta_D} \!
}
\right)^{\!2} \!=\! 0 ,
\end{array}
\label{CaseII}
\eeqa
whose solution is easily otained as\\[-14pt]
\beqa
\!\!\!\!
\begin{array}{l}
\mbox{arcsinh} \!
\left( \!\!
{\displaystyle
\frac{1}{x_{T}}
} \!\!
\right)
\!=\!
(0.283)^{\!-2} \!
\left\{ \! N(0) V \! \right\}^{\!-2} \!
\left\{ \! \pi \! N(0) \hbar \omega_D \! \right\} \!\!
\left( \!
{\displaystyle
\frac{T}{\theta_D}
} \!
\right) \!\!
\left[ \!\!\! {}^{{}^{{}^{{}^{{}^{{}^{{^{.}}}}}}}}
0.391 \!
\left\{ \! \pi \! N(0) \hbar \omega_D \! \right\} \!\!
\left( \!
{\displaystyle
\frac{T}{\theta_D}
} \!
\right)
\right. \\
\\[-12pt]
-
\left\{
(0.391)^2 \!
\left\{ \! \pi \! N(0) \hbar \omega_D \! \right\}^{\!2} \!
\left( \!
{\displaystyle
\frac{T}{\theta_D}
} \!
\right)^{\!2}
\!+
(0.283)^{\!2} \!
\left\{ \! N(0) V \! \right\}^{\!2} \!
0.782
\right. \\
\\[-12pt]
\left.
\left.
~~~~~~~~~~~~~~~~~~~~~~~~~~~~~~~~~
\!\times\!
\left\{
\ln (0.283)
\!-\!
\ln
\left\{ \! 2 \pi \! N(0) \hbar \omega_D \! \right\}
\!-\!
\ln
\left( \!
{\displaystyle
\frac{T}{\theta_D}
} \!
\right) \!
\right\} \!
\!\!\! {}^{{}^{{}^{{}^{{}^{{}^{.}}}}}}
\right\}^{\frac{1}{2}}
\right] \! ,
\end{array}
\label{CaseII2}
\eeqa
from which,
using again the approximate relation
$
\mbox{arcsinh} \!
\left( \!
1 / x_{T} \!
\right)
\!\approx\!
\left( x_{T} \!-\! 2 \right) \! / 2
$,
finally we have a solution for
$x_T~(N(0) V \!=\! 0.25,~N(0) \hbar \omega_D \!=\! 0.01~\mbox{and}~
\theta_{D} \!=\! 700\mbox{K})$
as\\[-8pt]
\beqa
\!\!\!\!
\begin{array}{l}
x_T
\!=\!
2
\!+\!
2~\!(0.283)^{-2}
(0.25)^{-2} \!
\left\{ 3.14 \!\times\! 0.01 \right\} \!
\left( \!
{\displaystyle
\frac{T}{700}
} \!
\right) \!
\left[ \!\! {}^{{}^{{}^{{}^{{}^{{}^{{^{.}}}}}}}}
0.391
\left\{ 3.14 \!\times\! 0.01 \right\} \!
\left( \!
{\displaystyle
\frac{T}{700}
} \!
\right)
\right. \\
\\[-14pt]
-
\left\{
(0.391)^2
\left\{ 3.14 \!\times\! 0.01 \right\}^{\!2} \!
\left( \!
{\displaystyle
\frac{T}{700}
} \!
\right)^{\!2}
\!+
(0.283)^{2}~\!
(0.25)^{2}
\!\times\!
0.782
\right. \\
\\[-14pt]
\left.
\left.
~~~~~~~~~~~~~~~~~~~~~~~~~~~~
\!\times\!
\left\{
\ln (0.283)
\!-\!
\ln
\left\{ 2 \!\times\! 3.14 \!\times\! 0.01 \right\}
\!-\!
\ln \!
\left( \!
{\displaystyle
\frac{T}{700}
} \!
\right) \!
\right\} \!
\!\!\! {}^{{}^{{}^{{}^{{}^{{}^{.}}}}}}
\right\}^{\frac{1}{2}}
\right] \! .
\end{array}
\label{solCaseII2}
\eeqa

It is very interesting to investigate behaviour of
the temperature dependence of the gap $\Delta_T$
for Case I and Case II.

\newpage


\def\thesection{\arabic{section}}
\setcounter{equation}{0}
\renewcommand{\theequation}{\arabic{section}.\arabic{equation}}

\section{Summary and further perspectives}

~~~We have concentrated 
on derivation of thermal {\em gap equations}
within the framework of Res-HBA.
From the Res-FB operators
${\cal F}_1$ and  ${\cal F}_2$
with {\em equal-gaps},
we have found the diagonalization conditions for them,
which are essentially of the same form as the former one.
It leads to the self-consistent 
thermal Res-HB gap equation
and makes possible to derive the new formulas to determine $T_c$
and the gaps near $T \!=\! 0$ and $T_c$.
The formula for Case I 
gives a {\it high} $T_c ^{\mbox{\scriptsize I}}$,
e.g.,
$T_c ^{\mbox{\scriptsize I}} \!=\! 72.87 $K
for $N(0)V \!=\! 0.25$ and 
Debye temperature $\theta _D \!=\! 700 $K.
This is in contrast with $T_c$
of the usual HB formula 
giving $T_c \!=\! 14.49 $K for the same values
of $N(0)V$ and $\theta _D$.
The formula for Case II
gives also a {\it high} $T_c ^{\mbox{\scriptsize II}}$,
e.g.,
$T_c ^{\mbox{\scriptsize II}} \!=\! 198 $K for the same values
of $N(0)V$ and $\theta _D$.
The temperature dependence of
the gap near $T \!=\! 0$ and $T_c$
becomes more complicated than that of
the usual HB and 
Abrikosov descriptions
\cite{AGD.65,Abrikosov.88}.
At intermediate temperature,
we have got the solutions of $\Delta_T$
for Case I and Case II.
We have taken
$[\det z_{12}]_T ^{1/2}
\!\!\approx\!
0.3
$ (Case I)
and
$N \! (0) \hbar \omega_{\!D} \!=\! 0.01$ (Case II)
to get real solutions.
We have got, however, 
$x_T \!=\! 0.65$ (Case I) and 1.05 (Case II)
for $T \!=\! 70$K
which are a little bit large compared with the real solutions.
Improvement of such results should be made.
Further, it would have been better to draw numerical aspects of the temperature dependence in all the correlation regimes.
This is possible in the near future.

For {\em unequal two-gaps},
it is also possible to realize 
the above-mentioned diagonalization condition
for Res-FB operators
${\cal F}_{rp}~(r \!=\! 1, 2)$.
Transforming by a unitary matrix
$\widehat{g}_{rp}$,
${\cal F}_{rp}$
is easily diagonalized.
Noticing the same correspondence as the one in
(\ref{equivalenceforcosandsin}),
$\cos \theta _{rp} \!\!\Rightarrow\!\! \cos \widehat{\theta }_{rp}$
and
$\sin \theta _{rp} \!\!\Rightarrow\!\! \sin \widehat{\theta }_{rp}$,
we assume each diagonalization condition 
(\ref{equivalenceforsin3})
holds even in this case.
Then we obtain coupled equations through
a function of $\Delta_{1T}$ and $ \Delta_{2T}$
expressed as\\[-12pt]
\beqa
\begin{array}{c}
1
\!=\!
{\displaystyle
\frac{
{\displaystyle
\frac{\varepsilon _p ^2}
{(\varepsilon _p ^2 \!+\! \Delta ^2 _{r T}) ^{3/2}}
} \!\!
\left( \!\!
{\displaystyle
- \frac{2{\cal F}^{\uparrow }_{r \Delta _{T}}}{\Delta _{r T}}
} \!\!
\right) \!\!
\left( \!
1 \!-\! 2 \widetilde{w}^{\uparrow }_{r p}
\right)
}
{
{\displaystyle
\frac{\varepsilon _p}
{(\varepsilon _p ^2 \!+\! \Delta ^2 _{r T}) ^{3/2}}
} \!
\left( \!
{\cal F}^{\uparrow }_{+ r \varepsilon p}
\!+\!
{\cal F}^{\uparrow }_{- r \varepsilon p} \!
\right) \!
\left( \!
1 \!-\! 2 \widetilde{w}^{\uparrow }_{r p}
\right)
}
}, ~
\widetilde{w}^{\uparrow }_{rp}
\!=\!
{\displaystyle
\frac{1}{1 \!+\!
e^{{\displaystyle \beta \widetilde{\epsilon }_{rp}}}}
}
\end{array}
\label{SCFconditionfortwogaps}
\eeqa\\[-6pt]
which reduces to
equation in R.H.S. of
(\ref{equivalenceforcosandsin})
if $\Delta_{1T} \!=\! \Delta_{2T}$.
The quantities
${\cal F}^{\uparrow }_{r, \Delta _{T}}$
and
${\cal F}^{\uparrow }_{r, \pm \varepsilon p}$
are given by the equations similar to
(5.9) in I
but with more complicated forms
of $\Delta_{1T}$ and $ \Delta_{2T}$.
For the time being,
as was done in the previous section
we here also use the function
$(\varepsilon _p ^2 \!+\! \Delta ^2 _{r T}) ^{3/2}$
by which we divide numerator and denominator, respectively,
in
(\ref{SCFconditionfortwogaps}).
After equating the numerator to the denominator and
using the relation
$
1 \!-\! 2 \widetilde{w}^{\uparrow }_{r p}
\!=\!
{\displaystyle
\tanh \!
\left(
\widetilde{\epsilon }_r / 2k_B T
\right)
}
$,
we sum up over $p$,
namely
integrate both sides of the equation over $\varepsilon$,
to achieve the optimized conditions.
We obtain coupled thermal
Res-HB gap equations
and
reach our ultimate goal of computing
temperature-dependent two-gaps.
Along such a strategy and method,
we will make a numerical analysis to demonstrate
the behaviour of temperature-dependent two-gaps.

To solve such a problem,
we must provide a rigorous thermal Res-HBA.
We have an expression for partition function in an
$SO(\!2N\!)$ CS rep $| g \rangle$
\cite{Perelomov.72},
$\mbox{Tr}(\!e^{-\beta H}\!)
\!=\!
2^{N \!-\! 1} \!\int \langle g |e^{-\beta H}| g \rangle dg$
($ \int \! dg$ is the group integration on group $SO(\!2N\!)$).
Following Fukutome
\cite{Fuku.88},
introducing the projection operator $P$ to the Res-HB subspace,
the partition function in the Res-HB subspace is computed as
$\mbox{Tr}(\!Pe^{-\beta H}\!)$.
This kind of trace formula is calculated within the Res-HB subspace
by using the Laplace transform of 
$e^{-\beta H}$
and the projection method
which leads to an infinite matrix continued fraction.
A thermal variation of 
the Res-HB free energy
is carried out after the inverse Laplace transform
of matrix elements of
$e^{-\beta H}$.
This is made parallel to the usual
thermal BCS theory,
which will be given in a separate paper.

\newpage


\begin{center}
{\bf Acknowledgements}
\end{center}
S. N. would like to
express his sincere thanks to
Professor Manuel Fiolhais for kind and
warm hospitality extended to
him at the Centro de F\'\i sica Computacional,
Universidade de Coimbra, Portugal.
This work was supported by FCT (Portugal) under the project
CERN/FP/83505/2008.
The authors are indebted to the former Professor
M. Ozaki of Kochi University
and
Professor N. Tomita of Yamagata University
for their invaluable discussions and useful comments.

\newpage


\vspace{1.0cm}
\leftline{\large{\bf Appendix}}
\appendix

\vspace{0.1cm}



\def\thesection{\Alph{section}}
\setcounter{equation}{0}
\renewcommand{\theequation}{\Alph{section}.\arabic{equation}}
\section{Resonating mean-field free energy
$F[{\cal Z}]_{\mbox{{\scriptsize Res}}}$}


~~~~Suppose  
a matrix ${\cal Z}$ to be a usual FB operator.
introduce a quadratic HB Hamiltonian and the usual HB free energy
\cite{Ozaki.85}\\[-24pt]
\beqa
\left.
\begin{array}{cc}
&
H[{\cal Z}] 
\equiv
{\displaystyle \frac{1}{2}}
[c^{\dagger },~c]~{\cal Z}
\left[
\begin{array}{c}
c ,\\
\\[-12pt]
c^{\dagger }
\end{array}
\right]
,~~{\cal Z}^{\dagger } = {\cal Z}, \\
\\[-16pt]
&
F[{\cal Z}] = \mbox{Tr}(\stackrel{\circ }{W}\! [{\cal Z}] H) 
+ \frac{1}{\beta }
\mbox{Tr}
\{
\stackrel{\circ }{W}\! [{\cal Z}] 
\ln (\stackrel{\circ }{W}\! [{\cal Z}])
\},~~
\stackrel{\circ }{W}\! [{\cal Z}] 
\!=\! 
{\displaystyle 
\frac{e^{-\beta H[{\cal Z}]}}
{\mbox{Tr}(e^{-\beta H[{\cal Z}]})}
} ,
\end{array}
\right\}
\label{usualHBfreeenergy}
\eeqa\\[-16pt]
which leads to\\[-28pt]
\beqa
\left.
\begin{array}{c}
F[{\cal Z}] 
= 
\langle H - H[{\cal Z}] \rangle _{{\cal Z}}
-{\displaystyle \frac{1}{\beta }} 
\ln \mbox{Tr} (e^{- \beta H[{\cal Z}]}) , \\
\\[-12pt]
\langle H \rangle _{{\cal Z}} 
\equiv
{\displaystyle \frac{\mbox{Tr}(e^{- \beta H[{\cal Z}]}H)}
{\mbox{Tr} (e^{- \beta H[{\cal Z}]})}}, ~~~
\langle H[{\cal Z}] \rangle _{{\cal Z}} 
\equiv 
{\displaystyle \frac{\mbox{Tr}(e^{- \beta H[{\cal Z}]}H[{\cal Z}])}
{\mbox{Tr} (e^{- \beta H[{\cal Z}]})}}.
\end{array}
\right\}
\label{usualHBfreeenergyandtraceform}
\eeqa
We have another form for this free energy,
i.e., a well-known formula 
expressed in terms of a usual HB density matrix
$W[{\cal Z}]$
as\\[-22pt]
\beqa
\left.
\begin{array}{c}
F[{\cal Z}] 
\!=\!
\langle H \rangle _{{\cal Z}} 
\!+\!
{\displaystyle \frac{1}{2}\frac{1}{\b}}
\mbox{Tr}
\left\{
W[{\cal Z}] \ln W[{\cal Z}] + (1_{2N} - W[{\cal Z}])
\ln (1_{2N} - W[{\cal Z}])
\right\},\\
\\[-10pt]
W[{\cal Z}] 
\equiv
\left[
\begin{matrix}
{R[{\cal Z}]&K[{\cal Z}]\\
\\[-6pt]
-K[{\cal Z}]^\ast &1_N - R[{\cal Z}]^\ast}
\end{matrix}
\right] ,
\end{array}
\right\}
\label{densityform}
\eeqa\\[-14pt]
and the trace formulas for the pair operators\\[-10pt]
\begin{equation}
\mbox{Tr}
\left\{
\stackrel{\circ }{W}\! [{\cal Z}]
\left(
{E^{\beta } }_{\alpha } + \frac{1}{2}
  {\delta }_{\beta \alpha }
\right)   
\right\}  
= R_{\alpha \beta } [{\cal Z}] ,~~~~~~
\mbox{Tr}
\left\{
\stackrel{\circ }{W}\! [{\cal Z}]
E_{\beta \alpha }
\right\}  
= K_{\alpha\beta}[{\cal Z}] .
\label{matrixelementsF1F2}
\end{equation}\\[-12pt]
We give ${\cal Z}$ by a direct sum of ${\cal F}_r$,
${\cal Z} = \sum_{r=1} ^n \oplus {\cal F}_r$
and assume each ${\cal F}_r(r \! = \! 1, \cdots , n)$ 
is a Res-FB operator.
Then, instead of 
the above $H[{\cal Z}]$ in
(\ref{usualHBfreeenergy}),
we introduce a quadratic Res-HB Hamiltonian\\[-30pt]
\beqa
\begin{array}{cc}
H[{\cal Z}]_{\mbox{{\scriptsize Res}}} 
\equiv &\!\!\!
{\displaystyle 
\frac{1}{2}}~
[c^{\dagger },~c,
\cdots,
c^{\dagger },~c,
\cdots,
c^{\dagger },~c] 
\left[ 
\begin{array}{ccccc}
{\cal F}_1&&&&\\
\\[-10pt]
&\ddots&&0&\\
&&{\cal F}_r&&\\
&0&&\ddots&\\
\\[-10pt]
&&&&{\cal F}_n
\end{array} 
\right] 
\left[
\begin{array}{c}
     c \\
     c^{\dagger }, \\[-8pt]
     \vdots \\[-8pt]
     c \\
     c^{\dagger }, \\[-8pt]
     \vdots \\[-8pt]
     c \\
     c^{\dagger }
\end{array}
\right] . 
\end{array}
\label{quadraticHBHamiltonian2}
\eeqa\\[-16pt]
Along the same way as the one in
(\ref{usualHBfreeenergy}),
the Res-HB free energy can also be defined.

Now let us introduce a projection operator
$P~(P^2 \!=\! P
\!=\!
P^\dagger)$ to
the Res-HB subspace,
$P |\Psi \rangle
\!=\!
|\Psi ^{\mbox{{\scriptsize Res}}} \rangle$
as
$
P 
\!\equiv\!
\sum _{r,s=1}^n |g _r \rangle (S^{-1})_{rs} \langle g _s | 
$,
where the
$|g _r \rangle$'s are HB wave functions and
$
S 
\!=\!
(S_{rs})
\left(
\!=\! 
[ \det z_{rs}]^{1 / 2}
\right)
$
is an $n \!\times\! n$ matrix composed of the overlap integrals
and $S^\dagger \!=\! S$.
Using the projection operator $P$,
we give the Res-HB free energy in the form\\[-14pt]
\beq
F[{\cal Z}]_{\mbox{{\scriptsize Res}}}
\! = \!
\mbox{Tr}(\stackrel{\circ}{W}\![{\cal Z}]_{\mbox{{\scriptsize Res}}} H)  
\! + \!
\frac{1}{\beta}
\mbox{Tr}
\left\{ \!
\stackrel{\circ}{W}\![{\cal Z}]_{\mbox{{\scriptsize Res}}} \ln
(\stackrel{\circ}{W}\![{\cal Z}]_{\mbox{{\scriptsize Res}}}) \!
\right\},
\stackrel{\circ}{W}\![{\cal Z}]_{\mbox{{\scriptsize Res}}} 
\! \equiv \!
\frac{Pe^{-\beta H[{\cal Z}]_{\mbox{{\scriptsize Res}}}}P}
{\mbox{Tr}(Pe^{-\beta H[{\cal Z}]_{\mbox{{\scriptsize Res}}}})} ,
\label{Res-HB free energy}
\eeq\\[-16pt]
in which,
by making Taylor expansion of
$\ln P \!=\! \ln \{1 - (1 - P) \}$
and using $P^2 \!=\! P$,
we have\\[-8pt]
\beq
\frac{1}{\beta} 
\mbox{Tr}
\left\{
\stackrel{\circ}{W}\![{\cal Z}]_{\mbox{{\scriptsize Res}}} \ln
(\stackrel{\circ}{W}\![{\cal Z}]_{\mbox{{\scriptsize Res}}})
\right\} 
\!=\!
- \langle H[{\cal Z}]_{\mbox{{\scriptsize Res}}} \rangle
_{{\cal Z};\mbox{{\scriptsize Res}}}
- \frac{1}{\beta} \ln \mbox{Tr} 
(Pe^{- \beta H[{\cal Z}]_{\mbox{{\scriptsize Res}}}}).
\label{traceform2}
\eeq
A natural extension of the HB free energy to
the Res-HB free energy is easily made.
Then, the Res-HB free energy is given as\\[-8pt]
\beq
F[{\cal Z}]_{\mbox{{\scriptsize Res}}}
    = \langle H - H[{\cal Z}]_{\mbox{{\scriptsize Res}}} \rangle
_{{\cal Z};\mbox{{\scriptsize Res}}}
- \frac{1}{\beta} \ln \mbox{Tr} 
(Pe^{- \beta H[{\cal Z}]_{\mbox{{\scriptsize Res}}}}),
\label{Res-HB free energy 2}
\eeq
\vspace{-0.5cm}
\beqa
\langle H[{\cal Z}]_{\mbox{{\scriptsize Res}}} \rangle
_{{\cal Z};\mbox{{\scriptsize Res}}} 
\equiv
{\displaystyle
\frac{\mbox{Tr}(Pe^{- \beta H[{\cal Z}]_{\mbox{{\scriptsize Res}}}}
H[{\cal Z}]_{\mbox{{\scriptsize Res}}})}
{\mbox{Tr} (Pe^{- \beta H[{\cal Z}]_{\mbox{{\scriptsize Res}}}})}
}, ~~
\langle H \rangle
_{{\cal Z};\mbox{{\scriptsize Res}}} 
\equiv
{\displaystyle
\frac{\mbox{Tr}(Pe^{- \beta H[{\cal Z}]_{\mbox{{\scriptsize Res}}}}H)}
{\mbox{Tr} (Pe^{- \beta H[{\cal Z}]_{\mbox{{\scriptsize Res}}}})}
} . 
\label{traceform3}
\eeqa

Consider the whole Res-HB subspace
$
|\Psi ^{\mbox{{\scriptsize Res}}(k)} \rangle
\!=\!
\sum _{t=1}^n
c_t ^{(k)}
|g _t \rangle , ~(k = 1, \cdots , n)
$
in which the Res-state with index
$k \!=\! 1$
and 
the Res-states with indices
$k \!=\! 2, \!\cdots\!, n$
stand for 
the Res-ground one and the Res-excited ones, respectively.
In order to determine 
the thermal $|g_r \rangle$'s and thermal mixing coefficients $c_r^{(k)}$'s
by the variational method,
we use a temperature dependent Lagrangian with
the Lagrange multiplier term $E$ to secure
the normalization condition 
$
\langle
\Psi ^{\mbox{{\scriptsize Res}}(k)} | 
\Psi  ^{\mbox{{\scriptsize Res}}(k)}
\rangle
=
1$:
\beqa
\begin{array}{c}
L[{\cal Z}]_{\mbox{{\scriptsize Res}}}^{\mbox{\scriptsize HB}} 
=
\sum_{k=1}^n
\sum_{r,s=1}^n
\left\{
H\left[
W_{rs}[{\cal Z}]_{\mbox{{\scriptsize Res}}}
\right] - E^{(k)}
\right\}
\cdot
\left[
\det z_{rs}
\right]^{\frac{1}{2}}
c_r ^{(k)*} c_s^{(k)} .
\end{array} 
\label{statisticalLagrangian}
\eeqa
The variation of 
(\ref{statisticalLagrangian}) 
is made in a quite parallel manner 
to the one in the previous works
\cite{Fuku.88,NishiFuku.91}.
We omit detaied derivations for such parallel cases.
From the variation of 
$L[{\cal Z}]_{\mbox{{\scriptsize Res}}}^{\mbox{\scriptsize HB}} $ 
with respect to $c^{(k) \ast}_{r}$ for any $k$, 
we get a thermal Res-HB CI equation 
to determine
$c^{(k)}_{s}$
\beqa
\begin{array}{c}
\sum_{s=1}^n 
\left\{ 
H\left[
W_{rs}[{\cal Z}]_{\mbox{{\scriptsize Res}}}
\right] - E^{(k)} 
\right\} 
\cdot [\det z_{rs}]^{\frac{1}{2}}c^{(k)}_{s} 
= 0 .
\end{array}
\label{statisticalResCIequation}
\eeqa
We define
$D_{rs}$ and $\tilde{D}_{rs}$ as
$D_{rs} 
\!\equiv\!
u_s z^{\!-1}_{rs} \delta u^\dagger _r$
and
$\tilde{D}_{rs} 
\!\equiv\!
\delta u_s z^{\!-1}_{rs} u^\dagger _r$,
respectively,
where
$
z_{rs}
\!\equiv\!
u^\dagger _r u_s
$
and
$
u_r ^{\mbox{{\scriptsize T}}}
\!\equiv\!
[b_r ^{\mbox{{\scriptsize T}}},a_r ^{\mbox{{\scriptsize T}}}]~
(a_r ~\mbox{and}~b_r \mbox{:r-th HB amplitudes})
$.
The variations of 
thermal HB interstate density matrix 
$W[{\cal Z}]_{\mbox{{\scriptsize Res}}}$  
and thermal overlap integral $[\det z]^{1 / 2}$ 
are given by
\beqa
\left.
\begin{array}{ll}
\delta W_{rs}[{\cal Z}]_{\mbox{{\scriptsize Res}}}
\!\!\!&
\!=\!
D_{rs}(1_{2N} \!-\! W_{rs}[{\cal Z}]_{\mbox{{\scriptsize Res}}})
\!+\!
(1_{2N} \!-\! W_{rs}[{\cal Z}]_{\mbox{{\scriptsize Res}}})
\tilde{D}_{rs} , \\
\\[-10pt]
\delta [\det z_{rs}]^{\frac{1}{2}}
\!\!\!&
\!=\!
{\displaystyle \frac{1}{2}} 
{\hbox{Tr}}(D_{rs} \!+\! \tilde{D}_{rs}) 
\cdot 
[\det z_{rs}]^{\frac{1}{2}} .
\end{array}
\right\}
\label{statisticaldeltaWanddetz}
\eeqa
We obtain also the variation of Hamiltonian matrix element 
$H[W_{rs} [{\cal Z}]_{\mbox{{\scriptsize Res}}}]$
as
\beqa
\left.
\begin{array}{cc}
\delta H[W_{rs} [{\cal Z}]_{\mbox{{\scriptsize Res}}}]
\!\!=\!\!
{\displaystyle \frac{1}{2}} {\hbox{Tr}} \!
\left\{ 
{\cal F}[W_{rs} [{\cal Z}]_{\mbox{{\scriptsize Res}}}]
\delta W_{rs} [{\cal Z}]_{\mbox{{\scriptsize Res}}} 
\right\} , ~\\
\\[-10pt]
{\cal F}[W_{rs}[{\cal Z}]_{\mbox{{\scriptsize Res}}}]
\!\!=\!\!
\left[
\begin{matrix}
{F_{rs}[{\cal Z}]_{\mbox{{\scriptsize Res}}}&
D_{rs}[{\cal Z}]_{\mbox{{\scriptsize Res}}} \\
\\[-6pt]
-D_{sr}^\ast [{\cal Z}]_{\mbox{{\scriptsize Res}}}&
-F_{sr}^\ast [{\cal Z}]_{\mbox{{\scriptsize Res}}}}
\end{matrix}
\right] , \\
\\[-6pt]
F_{rs;\alpha\beta}[{\cal Z}]_{\mbox{{\scriptsize Res}}} 
\equiv
{\displaystyle \frac{\delta H[W_{rs}
[{\cal Z}]_{\mbox{{\scriptsize Res}}}]}
{\delta R_{rs;\beta\alpha} 
[{\cal Z}]_{\mbox{{\scriptsize Res}}}}
= 
h_{\alpha\beta}+[\alpha\beta |\gamma\delta]
R_{rs;\delta\gamma}[{\cal Z}]_{\mbox{{\scriptsize Res}}}} ,\\
\\[-6pt]
D_{rs;\alpha\beta}[{\cal Z}]_{\mbox{{\scriptsize Res}}}
\equiv
{\displaystyle \frac{\delta H[W_{rs}
[{\cal Z}]_{\mbox{{\scriptsize Res}}}]}
{\delta K_{sr;\alpha\beta}^\ast 
[{\cal Z}]_{\mbox{{\scriptsize Res}}}}
= 
-\frac{1}{2}[\alpha\gamma |\beta\delta]
K_{rs;\delta\gamma}[{\cal Z}]_{\mbox{{\scriptsize Res}}}}.
\end{array}
\right\}
\label{statisticaldeltaH}
\eeqa
Following I and the Res-HF theory
\cite{Fuku.88},
writing
$
L[{\cal Z}]_{\mbox{{\scriptsize Res}}}^{\mbox{\scriptsize HB}} 
\! = \!
\sum_{k = 1} ^n
\sum_{r,s=1}^n
{\cal L}_{rs}^{\mbox{\scriptsize HB(k)}} 
[{\cal Z}]_{\mbox{{\scriptsize Res}}}
c_r ^{(k)*} c_s^{(k)}
$
and
$
{\cal L}_{rs}^{\mbox{\scriptsize HB(k)}}
[{\cal Z}]_{\mbox{{\scriptsize Res}}}
\! = \!
\{ H[W_{rs}[{\cal Z}]_{\mbox{{\scriptsize Res}}}]
\!-\!
E^{(k)} \}
\!\cdot\!
[\det z_{rs}]^{1 / 2}
$,
from the variation of
$L_{\mbox{{\scriptsize Res}}}^{\mbox{\scriptsize HB}}$
for any $k$,
we obtain a thermal Res-HB equation to determine
the thermal mean field wave function $u_r$ as
\beqa
\left.
\begin{array}{cc}
\sum_{k =1}^n
\sum_{s =1}^n {\cal K}_{rs}^{(k)} 
[{\cal Z}]_{\mbox{{\scriptsize Res}}}
c_r ^{(k)*} c_s^{(k)}
= 0, \\
\\[-6pt]
\!\!\!\!\!\!\!\!
{\cal K}_{rs}^{(k)}[{\cal Z}]_{\mbox{{\scriptsize Res}}} 
\! \equiv \!\!
\left\{ \!
(1_{\!2N} \!\! - \!\! W_{rs}[{\cal Z}]_{\mbox{{\scriptsize Res}}}) 
{\cal F} [W_{rs}[{\cal Z}]_{\mbox{{\scriptsize Res}}}] 
\!\! + \!\!
H[W_{rs}[{\cal Z}]_{\mbox{{\scriptsize Res}}}] 
\!\! - \!\! E^{(k)} \!
\right\}
\! \cdot \!
W_{rs}[{\cal Z}]_{\mbox{{\scriptsize Res}}} 
\! \cdot \!
[\det z_{rs}]^{\frac{1}{2}} .
\end{array} \!\!\!
\right \}
\label{thermalRes-HBequation}
\eeqa
From
(\ref{thermalRes-HBequation})
we can derive a thermal Res-HB eigenvalue equation.
See next Appendix.

\newpage



\def\thesection{\Alph{section}}
\setcounter{equation}{0}
\renewcommand{\theequation}{\Alph{section}.\arabic{equation}}
\section{Variation of resonating mean-field free energy 
$F[{\cal Z}]_{\mbox{{\scriptsize Res}}}^{\mbox{\scriptsize HB}}$}


~~~~
The thermal Res-HB coupled eigenvalue equations is expressed as follows:
\\[-14pt]
\beqa
\left.
\begin{array}{rl}
&
[{\cal F}_r [{\cal Z}]_{\mbox{{\scriptsize Res}}} u_r]_i 
=
\epsilon_{ri}u_{ri},~~
\epsilon_{ri}
\equiv 
\widetilde{\epsilon_r}_s
- \sum_{k =1}^n
\left\{ 
H[W_{rr}[{\cal Z}]_{\mbox{{\scriptsize Res}}}] - E^{(k)}
\right\} 
|c_r ^{(k)}|^2 , \\
\\[-4pt]
&
{\cal F}_r [{\cal Z}]_{\mbox{{\scriptsize Res}}}
\!=\!
{\cal F}_r [{\cal Z}]_{\mbox{{\scriptsize Res}}}^\dagger
\equiv
{\cal F}[W_{rr}[{\cal Z}]_{\mbox{{\scriptsize Res}}}]
\sum_{k =1}^n |c_r ^{(k)}|^2 \\
\\[-4pt]
&
~~~~~~~~~~~~~~~~~~~~~~~~~~~~~
+\!
\sum_{k =1}^n \! \sum_{s=1}^{\prime~n} 
\left\{
{\cal K}_{rs}^{(k)} [{\cal Z}]_{\mbox{{\scriptsize Res}}}
c_r ^{(k)*} c_s^{(k)}
+ 
{\cal K}_{rs}^{(k)\dagger} [{\cal Z}]_{\mbox{{\scriptsize Res}}}
c_r ^{(k)} c_s^{(k)*} 
\right\}  . 
\end{array} 
\right \}
\label{thermalRes-HBeigenvalueequation}
\eeqa
We call the $2N \!\times\! 2N$ matrix
${\cal F}_r [{\cal Z}]_{\mbox{{\scriptsize Res}}}$
the thermal Res-FB operator.
From now let us denote
$W_{rs}[{\cal Z}]_{\mbox{{\scriptsize Res}}},
{\cal F}_r [{\cal Z}]_{\mbox{{\scriptsize Res}}}$
and
${\cal K}_{rs}^{(k)}[{\cal Z}]_{\mbox{{\scriptsize Res}}}$
simply as
$W_{rs},{\cal F}_r$ and ${\cal K}_{rs}^{(k)}$,
respectively.
First, due to idempotent-like product properties
$W_{rs}W_{rr} \!\!=\!\! W_{rs}$ and $W_{sr}W_{rr} \!\!=\!\! W_{rr}$,
we have important relations
${\cal K}_{rs}^{(k)}W_{rr}
\!\!=\!\!
{\cal K}_{rs}^{(k)}$
and
${\cal K}_{rs}^{(k)\dagger } W_{rr}
\!\!=\!\!
\left\{
H[W_{sr}] \!\!-\!\! E^{(k)}
\right\}
W_{rr}
\!\cdot\!
[\det z_{sr}]^{\frac{1}{2}}$.
Next multiplication of
the second equation in
(\ref{thermalRes-HBeigenvalueequation})
by $W_{rr}$ from the right yields\\[-14pt]
\beqa
\!\!\!
\begin{array}{rl}
{\cal F}_r W_{rr} 
= &\!\!\!
{\cal F}[W_{rr}] W_{rr} \sum_{k =1}^n |c_r ^{(k)}|^2 \\
\\[-4pt]
& ~~~~~+
\sum_{k =1}^n \sum_{s = 1} ^{\prime~n}
\left[
{\cal K}_{rs}^{(k)} c_r ^{(k)*} c_s^{(k)}
+
\left\{H[W_{sr}] - E^{(k)}\right\} W_{rr} 
\cdot 
[\det z_{sr}]^{\frac{1}{2}} c_r ^{(k)} c_s^{(k)*}
\right] \\
\\[-4pt]
= &\!\!\!
{\cal F}[W_{rr}] W_{rr} 
\sum_{k =1}^n |c_r ^{(k)}|^2
\!-\!
\sum_{k =1}^n {\cal K}_{rr}^{(k)} |c_r ^{(k)}| ^2
\!-\! 
\sum_{k =1}^n 
\left\{H[W_{rr}] - E^{(k)}\right\} W_{rr} |c_r ^{(k)}| ^2 \\
\\[-4pt]
& ~~~+
\sum_{k =1}^n \sum_{s = 1} ^n
\left[
{\cal K}_{rs}^{(k)} c_r ^{(k)*} c_s^{(k)}
+ 
\left\{H[W_{sr}] - E^{(k)}\right\} 
\cdot 
[\det z_{sr}]^{\frac{1}{2}} 
c_s^{(k)*} \cdot W_{rr} c_r ^{(k)}
\right] \! . 
\end{array}
\label{modificationofFW}
\eeqa
Using (\ref{statisticalResCIequation}),
the second term in the last line of R. H. S. of
(\ref{modificationofFW}) is vanished.
Substituting to ${\cal K}_{rr}^{(k)}$ its explicit form
obtained from
(\ref{thermalRes-HBequation}),
thus, ${\cal F}_r W_{rr}$ is cast into\\[-16pt]
\beqa
\begin{array}{rl}
&\!\!\!\!\!\!\!\!
{\cal F}_r W_{rr}
= 
{\cal F}[W_{rr}] W_{rr} \sum_{k =1}^n |c_r ^{(k)}|^2 \\
\\[-4pt]
&\!\!\!\!
-
\sum_{k =1}^n  
\left[
(1_{2N} \! - \! W_{rr}) {\cal F} [W_{rr}] 
\! + \!
2\left\{H[W_{rr}] \! - \! E^{(k)}\right\}
\right]
\cdot W_{rr} |c_r ^{(k)}|^2 
\! + \!
\sum_{k =1}^n \sum_{s = 1} ^n
{\cal K}_{rs}^{(k)} c_r ^{(k)*} c_s^{(k)} \\
\\[-4pt]
&\!\!\!\!\!\!\!\!
= 
W_{rr} {\cal F}[W_{rr}] W_{rr} \!
\sum_{k =1}^n \! |c_r ^{(k)}|^2
\! - \!
\sum_{k =1}^n \!
2 
\left\{ \! H[W_{rr}] \! - \! E^{(k)} \! \right\} \! 
W_{rr} |c_r ^{(k)}| ^2 
\! + \!
\sum_{k =1}^n \! \sum_{s = 1} ^n \!
{\cal K}_{rs}^{(k)} c_r ^{(k)*} c_s^{(k)} \!,
\end{array}
\label{modificationofFW2}
\eeqa
and taking hermitian conjugate of both sides of
(\ref{modificationofFW2}),
we have
$
{\cal F}_r W_{rr} \!\!=\!\! W_{rr}{\cal F}_r
$.
This means the two hermitian matrices
${\cal F}_r$
and
$W_{rr}$
have common eigenvectors to diagonalize them
so that it leads to 
(\ref{thermalRes-HBeigenvalueequation}).
Therefore, 
(\ref{thermalRes-HBeigenvalueequation})
and the relation
${\cal F}_r W_{rr} \!\!=\!\! W_{rr}{\cal F}_r$
are equivalent.
Due to the idempotency relation
$
W_{rr}^2
\!\!=\!\!
W_{rr}
$,
${\cal F}_r W_{rr} \!\!=\!\! W_{rr}{\cal F}_r$
is equivalent to
$
{\cal F}_r W_{rr} \!-\! W_{rr}{\cal F}_r W_{rr}
\!\!=\!\!
0
$.
Further mutiplying
(\ref{modificationofFW2}) 
by $W_{rr}$ from the left
and using the explicit form of ${\cal K}_{rs}^{(k)}$,
we obtain\\[-16pt]
\beqa
\begin{array}{rl}
&\!\!\!\!
W_{rr} {\cal F}_r W_{rr}
= W_{rr} ^2 {\cal F}[W_{rr}] W_{rr} \sum_{k =1}^n |c_r ^{(k)}|^2
- \sum_{k =1}^n 
2 \left\{H[W_{rr}] - E^{(k)}\right\} W_{rr} ^2 |c_r ^{(k)}|^2 \\
\\[-4pt]
&~~+
\sum_{k =1}^n \sum_{s = 1} ^n
W_{rr}
\left\{
(1_{2N} - W_{rs}) {\cal F} [W_{rs}] + H[W_{rs}] - E^{(k)}
\right\}
\cdot W_{rs} \cdot 
[\det z_{rs}]^{\frac{1}{2}} c_r ^{(k)*} c_s^{(k)} \\
\\[-4pt]
&\!\!\!\!
= 
W_{rr} {\cal F}[W_{rr}] W_{rr} \sum_{k =1}^n |c_r ^{(k)}|^2
- \sum_{k =1}^n 
2 \left\{H[W_{rr}] - E^{(k)}\right\} W_{rr} |c_r ^{(k)}| ^2 .
\end{array}
\label{modificationofFW3}
\eeqa
Subtracting
(\ref{modificationofFW3})
from
(\ref{modificationofFW2}),
it is easy to derive an equivalence relation\\[-16pt]
\beqa
\begin{array}{c}
\sum_{k =1}^n \sum_{s = 1} ^n
{\cal K}_{rs}^{(k)} c_r ^{(k)*} c_s^{(k)}
=
{\cal F}_r W_{rr} - W_{rr} {\cal F}_r W_{rr} .
\end{array}
\label{modificationofWFW}
\eeqa
Thus,
the equivalence of
(\ref{thermalRes-HBeigenvalueequation}) 
with 
(\ref{thermalRes-HBequation})
has been proved. 
The above equivalent relation
(\ref{modificationofWFW})
also makes a crucial role in the variation of the Res-HB free energy.

Let us introduce the following Res-HB free energy
$F[{\cal Z}]_{\mbox{{\scriptsize Res}}}^{\mbox{\scriptsize HB}}$
which is quite similar to 
(\ref{densityform})
but involves the HB interstate density matrix instead of 
the usual HB density matrix:
\beqa
\left.
\begin{array}{l}
F[{\cal Z}]_{\mbox{{\scriptsize Res}}}^{\mbox{\scriptsize HB(1)}} 
= 
\sum_{k =1}^n \sum_{r,s =1} ^n
\left\{
H[W_{rs}] - E^{(k)}
\right\}
\cdot [\det z_{rs}]^{\frac{1}{2}} c_r ^{(k)*} c_s^{(k)} , \\
\\[-4pt]
F[{\cal Z}]_{\mbox{{\scriptsize Res}}}^{\mbox{\scriptsize HB(2)}} 
= 
{\displaystyle \frac{1}{2}\frac{1}{\b}}
\sum_{r, s =1} ^n
\mbox{Tr}
\left\{
W_{rs}\ln W_{rs} + (1_{2N} - W_{rs})\ln (1_{2N} - W_{rs}) 
\right\} ,\\
\\[-4pt]
F[{\cal Z}]_{\mbox{{\scriptsize Res}}}^{\mbox{\scriptsize HB}} 
=
F[{\cal Z}]_{\mbox{{\scriptsize Res}}}^{\mbox{\scriptsize HB(1)}} 
+
F[{\cal Z}]_{\mbox{{\scriptsize Res}}}^{\mbox{\scriptsize HB(2)}} .
\end{array}
\right\}
\label{ModificationofFreeEnergy}
\eeqa
We are now in a stage to make
a variation of the Res-HB free energy.
Using the variational formulas
(\ref{statisticaldeltaWanddetz}),
the thermal Res-HB equation
(\ref{thermalRes-HBequation}),
the equivalence relation 
(\ref{modificationofWFW}) 
and
the commutablity $[{\cal F}_r , W_{rr}] = 0$,
it is made as follows:
\beqa
\begin{array}{rl}
\delta F[{\cal Z}]_{\mbox{{\scriptsize Res}}}^{\mbox{\scriptsize HB(1)}} 
&\!\!\!\!
=
\sum_{r=1}^n
{\displaystyle \frac{1}{2}}\mbox{Tr}
\left\{
\sum_{k =1}^n \sum_{s=1}^n 
{\cal K}_{rs}^{(k)} c_r ^{(k)*} c_s^{(k)} u_r \delta u_r ^{\dagger}
\right\} \\
\\[-6pt]
&~~~~~~~~~~~~~~~~~~~~~~~~~~~~~~~~~~~+
\sum_{r=1}^n
{\displaystyle \frac{1}{2}}\mbox{Tr}
\left\{
\delta u_r u_r ^{\dagger}
\sum_{k =1}^n \sum_{s=1}^n 
{\cal K}_{rs}^{(k)\dagger } c_r ^{(k)} c_s^{(k)*} 
\right\} \\
\\[-6pt]
&\!\!\!\!
= 
\sum_{r=1}^n
{\displaystyle \frac{1}{2}}\mbox{Tr}
\left\{
\left[
({\cal F}_r W_{rr} \! - \! W_{rr}{\cal F}_r W_{rr})
u_r \delta u_r ^{\dagger}
\right]
\! + \!
\left[
\delta u_r u_r ^{\dagger}
(W_{rr} {\cal F}_r \! - \! W_{rr}{\cal F}_r W_{rr})
\right]
\right\} \\
\\[-6pt]
&\!\!\!\!
= 
\sum_{r=1}^n
{\displaystyle \frac{1}{2}}\mbox{Tr}
\left[
{\cal F}_r (1_{2N} - W_{rr})\delta W_{rr}
\right] ,
\end{array}
\label{variationoffreeenergy1}
\eeqa\\[-28pt]
\beqa
\begin{array}{ll}
\!\!\!\!\!\!\!\!
\delta F[{\cal Z}]_{\mbox{{\scriptsize Res}}}^{\mbox{\scriptsize HB(2)}} 
&\!\!\!
=
{\displaystyle \frac{1}{2}\frac{1}{\b}}\sum_{r=1}^n \mbox{Tr}
\left[
\ln
\left\{
W_{rr}(1_{2N} - W_{rr})^{-1}
\right\}
(1_{2N} - W_{rr})
\delta W_{rr}
\right.\\
\\[-8pt]
&
\left.
~~~~~~~~
+
\sum_{s=1}^{\prime~n}
(1_{2N} - W_{rs})
\ln
\left\{
W_{rs}(1_{2N} - W_{rs})^{-1}
\right\}
W_{rs}
u_r \delta u_r ^{\dagger}
+
\mbox{(h.c.)}
\right] ,
\end{array}
\label{variationoffreeenergy2}
\eeqa
the last of which
has no contribution
since
$(1_{2N} \!-\! W_{rs}) W_{rs} \!=\! 0$.
Then, the variational equation 
$
\delta F[{\cal Z}]_{\mbox{{\scriptsize Res}}}^{\mbox{\scriptsize HB}} 
\!=\!
\delta F[{\cal Z}]_{\mbox{{\scriptsize Res}}}^{\mbox{\scriptsize HB(1)}} 
\!+\!
\delta F[{\cal Z}]_{\mbox{{\scriptsize Res}}}^{\mbox{\scriptsize HB(2)}} 
\!=\! 0
$
leads to
$
\ln
\left\{
W_{rr}(1_{2N} \!-\! W_{rr})^{-1}
\right\}
\!=\! 
- \b {\cal F}_r
$,
in which we have used the variational relations
$
\delta W_{rr}
\!=\!
u_r \delta u_r ^{\dagger } \!+\! \delta u_r u_r ^{\dagger }
$
and
$
\delta u_r ^{\dagger } u_r  \!+\! u_r ^{\dagger } \delta u_r
\!=\! 0
$.
Thus we obtain
$
W_{rr}(1_{2N} \!-\! W_{rr})^{-1}
\!=\! 
\exp \{- \b {\cal F}_r \}
$
from which we have
$
W_{rr} 
\!=\! 
\exp \{- \b {\cal F}_r \} (1_{2N} \!-\! W_{rr})
$.
Finally we can reach the $r$-th thermal HB density matrix
$W_{rr}[{\cal Z}]_{\mbox{{\scriptsize Res}}}$ 
expressed in terms of
the $r$-th thermal Res-FB operator 
${\cal F}_r [{\cal Z}]_{\mbox{{\scriptsize Res}}}$ as
$
W_{rr}[{\cal Z}]_{\mbox{{\scriptsize Res}}}
\!=\!
{\displaystyle 
\frac{1}
{1_{2N} \!+\! \exp 
\{ \b {\cal F}_r [{\cal Z}]_{\mbox{{\scriptsize Res}}}
\}
}
}
$
where we have used the relation
$[{\cal F}_r , W_{rr}] \!=\! 0$.
By using a Bogoliubov transformation $g_r$,
$
W_{rr}[{\cal Z}]_{\mbox{{\scriptsize Res}}}
$ 
is diagonalized as follows:\\[-24pt]
\beqa
\left.
\widetilde{W}_r 
\!=\!
r ^{\dagger } 
W_{rr}[{\cal Z}]_{\mbox{{\scriptsize Res}}}
g_r
\!=\!\!
\left[ \!\!
\begin{array}{cc}
\left\{\widetilde{w}_{ri}\right\} & 0\\
&\\[-6pt]
0 & 1_N - \left\{\widetilde{w}_{ri}\right\}
\end{array} \!\!
\right] , \!\!
\begin{array}{c}
\widetilde{w}_{ri}
\!=\!
{\displaystyle
\frac{1}{1 \!+\! \exp \{ \b \widetilde{\epsilon}_{ri} \}}
}
~(r \!=\! 1, \cdots, n) , \\
\\[-10pt]
1 \!-\! \widetilde{w}_{ri}
\!=\!
{\displaystyle
\frac{1}{1 \!+\! \exp \{-\b \widetilde{\epsilon}_{ri} \}}
}
~(i \!=\! 1, \cdots, N).
\end{array} \!\!
\right\}
\label{tildeCalFrmatrix}
\eeqa
The diagonalization of
the $r$-th thermal Res-FB operator 
${\cal F}_r [{\cal Z}]_{\mbox{{\scriptsize Res}}}$
by the same Bogoliubov transformation $g_r$
leads us to the eigenvalue
$\epsilon_{ri}$.
To this eigenvalue by adding a term
$\sum_{k=1} ^n
\left\{ 
H[W_{rr}[{\cal Z}]_{\mbox{{\scriptsize Res}}}] 
\! - \! E^{(k)}
\right\} 
|c_r ^{(k)} |^2 
\!\cdot\! 1_{2N}$,
the usual HB type of the eigenvalue
$\widetilde{\epsilon}_{ri}$
is realized.
Using 
(\ref{tildeCalFrmatrix})
we can derive
the inverse transformation of
(\ref{tildeCalFrmatrix})
in the form\\[-16pt]
\beqa
\begin{array}{rl}
g_r ^{\dagger} W_{rr}[{\cal F}_r ] g_r
&\!\!\!\! 
=
g_r ^{\dagger }
{\displaystyle
\frac{1}
{1_{2N} \! + \! \exp
\left[ 
\b ({\cal F}_r [{\cal Z}]_{\mbox{{\scriptsize Res}}} 
\! + \! \sum_{k=1} ^n 
\left\{ 
H[W_{rr}[{\cal Z}]_{\mbox{{\scriptsize Res}}}] 
\! - \! E^{(k)}
\right\} 
|c_r ^{(k)}|^2 
\cdot 
1_{2N}) 
\right]
}  g_r \!
} \\
\\[-10pt]
& 
=
\widetilde{W}_r 
=  \!
\left[ \!\!
\begin{array}{cc}
\left\{\widetilde{w}_{ri}\right\} & 0 \\
&\\[-8pt]
0 & 1_N \!-\! \left\{\widetilde{w}_{ri}\right\} \!\!
\end{array}
\right] .
\end{array}
\label{CalFrmatrix}
\eeqa

\newpage



\def\thesection{\Alph{section}}
\setcounter{equation}{0}
\renewcommand{\theequation}{\Alph{section}.\arabic{equation}}
\section{$\!\!$Calculations of $\sum _p \! A_p,\sum _p \! B_p$ and
$\sum _p \! C_p$ at intermediate temperature}


~~
First we give a integral formula\\[-10pt]
\beq
\int _{0} ^\infty
\!\!\! dy
\left\{
\frac
{1}{y ^3}
\tanh y
-\!
\frac{1}{y ^2}
\mbox{sech} ^2 y
\right\} 
\!=\!
\frac{7}{\pi ^2}\zeta (3) ,
\label{integformula}
\eeq
which can be derived
by using the famous mathematical formulas
\cite{GraRyz.63}\\[-18pt]
\beqa
\!\!\!\!
\begin{array}{c}
{\displaystyle
\frac{1}{y}
\tanh y
}
\!=\!
8 \! \sum_{m \!=\! 1}^{\infty } \!
{\displaystyle
\frac{1}{(2m \!\!-\!\! 1)^2 \pi ^2 \!\!+\!\! 4 y ^2}
} , ~
{\displaystyle
\frac{1}{y ^3}
\tanh y
-\!
\frac{1}{y ^2}
\mbox{sech} ^2 y
}
\!=\!
64 \! \sum_{m \!=\! 1}^{\infty } \!
{\displaystyle
\frac{1}
{
\left\{ \!
(2m \!\!-\!\! 1)^2 \pi ^2 \!\!+\!\! 4 y ^2 \!
\right\} ^2
}
}.
\end{array}
\label{mathformulas}
\eeqa
Adopting a new integral variable
$
y
\!=\!
(2m \!\!-\!\! 1) \pi /2
\!\cdot\!
\tan \theta
$,
an integral of the second formula in
(\ref{mathformulas}) is easily carried out for
$\hbar \omega_D \gg 1$
as\\[-16pt]
\beqa
\begin{array}{c}
{\displaystyle
64 \!
\int _{0} ^{y_T ^{\mbox{\scriptsize II}}\rightarrow \infty }
}
\!\!\! d{y} \!
\sum _{m \!=\! 1}^{\infty } \!
{\displaystyle
\frac{1}
{
\left\{ \!
(2m \!\!-\!\! 1)^2 \pi ^2 \!\!+\!\! 4 y ^2 \!
\right\} ^2
}
}
\!=\!
32
\sum _{m \!=\! 1}^{\infty } \!
{\displaystyle
\frac{1}
{(2m \!\!-\!\! 1) ^3 \pi ^3} \!
\int _{0} ^{\frac{\pi }{2}}
\!\!\! d\theta
\frac{1}{1 \!\!+\!\! \tan ^2 \theta }
\!\!=\!\!
\frac{7}{\pi ^2}\zeta (3)
},
\end{array}
\label{IntegralCalofA2}
\eeqa\\[-10pt]
where we have used
$\sum _{m = 1}^{\infty }
(2m \!\!-\!\! 1)^{-3}
\!=\! 
(7/8) \!\cdot\! \zeta (3)$,
and
$\zeta (3) 
\!=\!
\pi ^3 \!/ 25.79436
$
\cite{GraRyz.63}.

Next the modified QP energy $\widetilde{\varepsilon }$
is approximated as
$
\widetilde{\varepsilon } ^{(\pm)}
\!=\!
\sqrt{\varepsilon ^2 \!+\! \widetilde{\Delta }_T ^2}
\left\{ \!
2 \! \left( 1 \!\pm\! [\det z_{12}]_T ^{1/2} \right) \!
\right\}^{-1}
$.
Let us introduce a new variable $y$ by
$
\varepsilon
\!=\!
4 \!
\left(
1 \!\pm\! [\det z_{12}]_T ^{1/2}
\right) \!
k_B T y
$
and quantities
$
\widetilde{x}_T
\!=\!
\widetilde{\Delta }_T/\hbar \omega_D
$
and
$
y_T  ^{(\pm)}
\!\!=\!\!
\sqrt{\varepsilon ^2 \!+\! \widetilde{\Delta }_T ^2}
\left\{ \!
4 \! \left( 1 \!\pm\! [\det z_{12}]_T ^{1/2} \right) \!
\right\}^{-1} \!
\hbar \omega_D / k_B T
$
where
$
\widetilde{\Delta }_T
\!\equiv\!
\Delta _T N(0) V
\mbox{arcsinh} \!
\left(
\hbar \omega_D / \Delta_T
\right)
$.
If
$\varepsilon \!\gg\! \Delta_T$,
$\sum _p A_p,~\sum _p B_p$ and  $\sum _p C_p$ in
(\ref{DefinitionsofABandC})
are recast to the following integrals up to
$\widetilde{\Delta }_T$:\\[-10pt]
\beqa
\begin{array}{c}
{\displaystyle
\frac{\sum _p \! A_p}{2 N(0)}
}
\!\simeq\!
{\displaystyle
\int _{0} ^{\hbar \omega_D}
\!\!\!\! d \varepsilon 
\frac
{1}
{\sqrt{\varepsilon  ^2 \!+\! \widetilde{\Delta }_T ^2 }}
\tanh \!
\left( \!\!
\frac{\widetilde{\varepsilon } ^{(\pm)}}{2k_B T} \!\!
\right)
\!-\!\!
\int _{0} ^{\hbar \omega_D}
\!\!\!\! d \varepsilon
\frac
{ \widetilde{\Delta }_T ^2}
{\varepsilon ^2 \sqrt{\varepsilon ^2 \!+\! \widetilde{\Delta }_T ^2 }}
\tanh \!
\left( \!\!
\frac{\widetilde{\varepsilon } ^{(\pm)}}{2k_B T} \!\!
\right)
} \\
\\[-12pt]
\!\!\!\!\!\!
\!=\!
{\displaystyle
\int _{0} ^{y_T ^{(\pm)}}
\!\!\! dy
}
\left\{ \!
{\displaystyle
1
\!-\!
\frac{
\left( \!
y_T ^{(\pm)}
\widetilde{x}_T \!
\right) ^2}{y ^2}
} \!
\right\} \!
{\displaystyle
\frac
{1}
{
\sqrt{y  ^2 \!\!+\!\!
\left( \!
y_T ^{(\pm)}
\widetilde{x}_T \!
\right) ^2 }
}
} \!
\tanh \!
\left[ \!
\sqrt{y ^2 \!\!+\!\! 
\left( \!
y_T ^{(\pm)}
\widetilde{x}_T \!
\right) ^2 }
\right] ,
\end{array}
\label{CalofA}
\eeqa
\vspace{-0.3cm}
\beqa
\begin{array}{c}
\!\!\!\!
{\displaystyle
\frac{\widetilde{\Delta }_{\!T} \!\! \sum _p \! B_p}{N(0)}
}
\!\!\simeq\!\!\!
{\displaystyle
\int _{-\hbar \omega_D} ^{\hbar \omega_D} \!\!
\!\!\!\! d \varepsilon \!
}
{\displaystyle
\frac{\widetilde{\Delta }_T}
{\varepsilon \! \sqrt{\varepsilon ^2
\!\!+\!\!
\widetilde{\Delta }_T ^2}}
} \!
\tanh \!
\left( \!\!
{\displaystyle
\frac{\widetilde{\varepsilon } ^{(\pm)}}{2k_B T}
} \!\!
\right)
{\displaystyle
\!-
{\displaystyle \frac{1}{2}} \!\!
\int _{-\hbar \omega_D} ^{\hbar \omega_D} \!\!
\!\!\!\! d \varepsilon \!
}
{\displaystyle
\frac{\widetilde{\Delta }_T ^3}
{\varepsilon ^3 \! \sqrt{\varepsilon ^2
\!\!+\!\!
\widetilde{\Delta }_{\!T} ^2}}
} \!
\tanh \!
\left( \!\!
{\displaystyle
\frac{\widetilde{\varepsilon } ^{(\pm)}}{2k_B T}
} \!\!
\right)
\!=\! 0,
\end{array}
\label{CalofB}
\eeqa
\vspace{-0.3cm}
\beqa
\!\!\!\!\!\!
\begin{array}{rl}
&
{\displaystyle
\frac{ \widetilde{\Delta }_T ^2 \! \sum _p \! C_p}{2 N(0)}
}
\!\!\simeq\!\!
{\displaystyle
\int _{0} ^{\hbar \omega_D}
\!\!\!\!\! d \varepsilon
}
{\displaystyle
\frac{\widetilde{\Delta }_T ^2}
{\varepsilon ^2 \! \sqrt{\varepsilon  ^2 \!\! + \!\! \widetilde{\Delta }_T ^2}} \!
\tanh \!
\left( \!\!
\frac{\widetilde{\varepsilon }^{(\pm)}}{2k_B T} \!\!
\right)
\!-\!\!
\int _{0} ^{\hbar \omega_D}
\!\!\!\!\! d \varepsilon
\frac
{ \widetilde{\Delta }_T ^4}
{\varepsilon  ^4 \! \sqrt{\varepsilon ^2 \!\! + \!\! \widetilde{\Delta }_T ^2}} \!
\tanh \!
\left( \!\!
\frac{\widetilde{\varepsilon }^{(\pm)}}{2k_B T} \!\!
\right)
}\\
\\[-12pt]
&=
{\displaystyle
\left( \!
y_T ^{(\pm)}
\widetilde{x}_T \!
\right) ^{\!2} \!
\int _{0} ^{y_T ^{\mbox{\scriptsize II}}}
\!\!\! dy \!
}
\left\{ \!
{\displaystyle
1
\! - \!
\frac{\left(
y_T ^{(\pm)}
\widetilde{x}_T \!
\right) ^{\!2}}{y ^2}
} \!
\right\}
{\displaystyle
\frac{1}
{y ^2 \! \sqrt{y  ^2 \!\! + \!\!
\left(
y_T ^{(\pm)}
\widetilde{x}_T \!
\right) ^{\!2} }
}
}
\tanh \!
\left[ \!
\sqrt{y  ^2 \!\! + \!\!
\left(
y_T ^{(\pm)}
\widetilde{x}_T \!
\right) ^{\!2} }
\right] .
\end{array}
\label{CalofC}
\eeqa
To get a finite value of
$\sum _p \! A_{p}$,
expanding
(\ref{CalofA})
around
$
y_{T} ^{\!(\!\pm\!)\!}
\widetilde{x}_{\!T} \! ,
$
($\!$\ref{CalofA})
is boldly
approximated as\\[-14pt]
\beqa
\begin{array}{c}
{\displaystyle
\frac{\sum _p A_p}{2 N(0)}
}
\!\simeq\!
{\displaystyle
\int _{0} ^{y_T ^{(\pm)}}
\!\!\! dy
\frac{1}{y}
\tanh y
\!-\!
\frac{3}{2}
\left( \!
y_T ^{(\pm)}
\widetilde{x}_T \!
\right) ^2 \!\!
\int _{0} ^{y_T ^{(\pm)}}
\!\!\! dy
\left\{ \!
\frac
{1}{y ^3}
\tanh y
\!-\!
\frac{1}{y ^2}
\mbox{sech} ^2 y \!
\right\}
}.
\end{array}
\label{CalofA2}
\eeqa
In a similar way 
we also get a roughly approximated integral form for 
(\ref{CalofC}) as\\[-14pt]
\beqa
\begin{array}{c}
{\displaystyle
\frac{\widetilde{\Delta }_T ^2 \!\sum _p \! C_p}{2 N(0)}
}
\!\simeq\! 
{\displaystyle
\left( \!
y_T ^{(\pm)}
\widetilde{x}_T \!
\right) ^2 \!\!
\int _{0} ^{y_T ^{(\pm)}}
\!\!\! dy \!
\left\{ \!
\frac
{1}{y ^3}
\tanh y
\!-\!
\frac{1}{y ^2}
\mbox{sech} ^2 y \!
\right\}
} .
\end{array}
\label{CalofC2}
\eeqa
Integrations of
(\ref{CalofA2}) 
and 
(\ref{CalofC2})
are easily made by using the integration formula
(\ref{integformula})
if we take
the upper-value
$y_T ^{(\pm)}$
to be infinite.

\newpage


\end{document}